\documentclass[12pt]{article}
\pdfoutput=1        
    
\usepackage{hyperref}     
\usepackage{amsmath}       
\usepackage{amssymb} 
\usepackage{graphicx} 
\usepackage{url}
\usepackage{enumerate}
\usepackage{color}
\usepackage{ulem}
 
\usepackage{float,wrapfig,color} 

\allowdisplaybreaks[1]
\def\eq#1{(\ref{#1})}
\def\s[#1\s]{\begin{align}\begin{split}#1\end{split}\end{align}}
\def\[#1\]{\begin{align}#1\end{align}}

\def\bpsi{{\bar \psi}}
\def\tJ{{\tilde J}}
\def\K{K}
\def\tK{{\tilde \K}}
\def\G{G}
\def\tG{{\tilde \G}}
\def\tv{{\tilde v}}
\def\vp{v^\perp}
\def\sdet{{{\rm s}\hspace{-.07cm}\det}}
\def\dels{\delta^{\rm susy}}
\def\pPhi{\Phi^\perp}
\def\abs2#1{\left| #1 \right|^2}
\def\pvp{\Phi^{\perp}}
\def\pvpc{\Phi^{{\rm c}\perp}}
\def\ql#1{Q^{(#1)}}

\def\gc{\gamma}
\def\tgc{\tilde \gamma}

\def\ss{{[\hspace{-0.06cm}[}}
\def\se{]\hspace{-0.06cm}]}

\def\gt#1#2{g^{(1,p-1)}_{#1\, #2}}
\def\gtp#1#2{g^{(\perp,p-1)}_{#1\, #2}}
\def\gtinv#1#2{g^{(1,p-1) -1}_{#1\, #2}}
\def\bgtinv#1#2{\bar g^{(1,p-1) -1}_{#1\, #2}}
\def\gq#1#2#3{g^{(#1)}_{#2#3}}
\def\gqinv#1#2#3{g^{(#1)\,-1}_{#2#3}}
\def\bgq#1#2#3{\bar g^{(#1)}_{#2#3}}
\def\gtpinv#1#2{g^{(\perp,p-1)-1}_{#1\, #2}}

\def\Ipara{I^{(1,p-1)}}
\def\Iperp{I^{(1,p-1)}_\perp}
\def\tIpara{I^{(\perp,p-1)}}
\def\tIperp{I^{(\cdot,p-1)}_\perp}

\def\hm{{\hat M}}
\def\hp{{\hat \Phi}}
\def\K{K}
\def\gtwo#1#2#3#4{g^{(#1,#2)}_{#3\,#4}}
\def\gthree#1#2#3#4#5{g^{(#1,#2,#3)}_{#4\,#5}}
\def\A{A}
\def\B{B}
\def\e{e}
\def\ep{e^\Phi}
\def\tP{{\tilde \Phi}}
\def\tK{{\tilde \K}}
\def\tM{{\tilde M}}
\def\tJ{{\tilde J}}

\def\re{{\rm Re}}
\def\im{{\rm Im}}

\def\bk{\bar K}

\def\svec#1#2{\genfrac{(}{)}{0pt}{}{#1}{#2}}

\def\rhocL{\rho^{\rm c}_L}
\def\rhoc{\rho^{\rm c}}
\def\Jc{J^{\rm c}}
\def\Phic{\Phi^{\rm c}}
\def\Sc{S^{\rm c}}
\def\ZcL{Z^{\rm c}_L}

\def\GPPc{\G_{\kappa\kappa'} \Phic_{l\kappa}\Phic_{l\kappa'}}
\def\GPPcd{\G_{\kappa''\kappa'''} \Phic_{l'\kappa''}\Phic_{l'\kappa'''}}
\def\mat2#1#2{
\left( 
\begin{matrix}
#1 \\ #2
\end{matrix}
\right)}
\def\Hc{H^{\rm c}}

\begin{document}

\begin{titlepage} 

\title{
\hfill\parbox{4cm}{ \normalsize YITP-26-47}\\   
\vspace{1cm} 
Joint distributions of eigenvectors\\
of symmetric random tensors
}

\author{Naoki Sasakura\footnote{sasakura@yukawa.kyoto-u.ac.jp}
\\
{\small{\it Yukawa Institute for Theoretical Physics, Kyoto University, }}\\
{\small{\it Kitashirakawa, Sakyo-ku, Kyoto 606-8502, Japan}}\\
{\small {\it and } } \\
{\small{\it CGPQI, Yukawa Institute for Theoretical Physics, Kyoto University,}} \\
{\small{\it Kitashirakawa, Sakyo-ku, Kyoto 606-8502, Japan}}
}
 
\date{\today} 
 

\maketitle 

\begin{abstract}
We compute the joint distributions of arbitrary numbers of eigenvectors
of real and complex symmetric random tensors by the quantum field theoretical methods 
which were previously used to compute the mean distributions.
We obtain the random matrix representations 
and the large-dimension asymptotics of the joint distributions. 
The latter can be expressed by a common function of tensor geometries, 
extending the universality found for the mean distributions to the joint distributions.  
Several crosschecks of our results are carried out by Monte Carlo computations. 
\end{abstract}
\end{titlepage}  
 
\section{Introduction}
\label{sec:introduction}
Since random matrix theory was first used to simulate the spectra of nuclei \cite{Wigner} by eigenvalues
of random matrices,
it has been applied to various problems in wide areas \cite{Brezin:1977sv,matrix}.  
Random tensor theory \cite{gurau,tanasa} has also been widening its areas of applications,  
since it was first introduced in the context of quantum 
gravity \cite{Ambjorn:1990ge,Sasakura:1990fs,Godfrey:1990dt}. 
In fact, while matrix eigenvalues/vectors ubiquitously appear in linear problems, so do
tensor eigenvalues/vectors \cite{Qi,lim,cart} in various non-linear problems.
For instance, they appear in spin glasses \cite{pspin,pedestrians,randommat},
quantum information theory \cite{shi,barnum}, 
industrial mathematics \cite{qibook}, gravitational systems \cite{Evnin:2021buq}, 
artificial intelligence \cite{Ouerfelli:2022rus},
string theory \cite{Biggs:2023mfn}, classical/quantum chaos \cite{Bianchi:2025kna}, and so on.

In spite of the growing importance of tensor eigenvalues/vectors, 
however, their properties are not yet thoroughly known because of their intricacies: 
numbers of eigenvalues are massive, generally exponential in dimensions of tensors \cite{cart}, 
and computing all eigenvalues/vectors of arbitrarily given tensors is extremely hard \cite{nphard}.
Considering the intricacies, it would be fruitful to study eigenvalues/vectors for random tensors,
since the problems would become more tractable by considering averaged properties 
over random tensors.
In fact earlier works \cite{randommat,fyodorov1,secondmoment,realnum1,realnum2, Evnin:2020ddw,Gurau:2020ehg} provided fruitful results in this line.

One way to capture the properties of eigenvalues/vectors for random tensors is to study their distributions. 
Some earlier works discussed such distributions in the context of the $p$-spin spherical
spin glass model \cite{pspin,pedestrians}\footnote{In the context of the spherical $p$-spin model, 
the critical points of the energy landscape correspond to the eigenvectors.}; the mean distributions of 
the eigenvalues of the real \cite{randommat} and 
complex \cite{Kent-Dobias:2020egr,Kent-Dobias2} symmetric random tensors, and the joint distributions 
of up to three eigenvectors of the real symmetric random tensor \cite{secondmoment,Parisi,jointthree}.
Recently, the kinds of random tensors have been
extended for the mean distributions
\cite{Sasakura:2024awt,Dartois:2024zuc,Majumder:2024ntn,Delporte:2025tjp,Dartois:2025ggp}.
Part of the results provided the large-dimension\footnote{Dimension here denotes dimensions of tenors.} 
asymptotics of the geometric measure of quantum entanglement \cite{shi,barnum} for the first time, 
which agreed with earlier numerical results \cite{Fitter:2022vpr}.

In this paper, we study the joint distributions of arbitrary numbers of the eigenvectors of the real 
and complex symmetric random tensors. 
We use the quantum field theoretical methods which were successfully applied to compute the mean 
distributions of the eigenvalues/vectors of various types of random tensors 
\cite{Sasakura:2022zwc,Sasakura:2022iqd,Sasakura:2022axo,Sasakura:2023crd,Delporte:2024izt,
Sasakura:2024awt,Majumder:2024ntn,Delporte:2025tjp}. 
We obtain the random matrix representations and the large-dimension asymptotics of the joint distributions.
Joint distributions describe mutual probabilistic correlations among  eigenvalues/vectors, 
and have important applications, such as proving that fluctuations of distributions vanish in asymptotic
limits \cite{secondmoment} and studying detailed structures of energy landscapes \cite{Parisi,jointthree}.
Moreover, in view of various universalities holding in the eigenvalue distributions of random 
matrices \cite{matrixuniv}, it would also be interesting to pursue parallel matters in random tensors.
In fact a universality was found that the large-dimension asymptotics of the mean distributions of 
various types of random tensors are expressed by a single function \cite{Delporte:2025tjp}. 
In this paper we show that the large-dimension asymptotics of arbitrary numbers of the joint distributions 
of the real and complex symmetric random tensors can be expressed by a single function of
tensor geometric quantities.
Though the finding is just for these two kinds of random tensors, it would suggest 
a possibility of universality of joint distributions over various types of random tensors.

This paper is organized as follows. In Section~\ref{sec:rsmean} we reproduce the mean distribution
of the eigenvectors of the real symmetric random tensor. 
The results themselves are all overlapping with the previous studies \cite{pspin,randommat,Sasakura:2022iqd}, 
but some technical details used in later sections are described in this simplest setting. In particular, 
the technical procedures of obtaining the random matrix representations and the large-dimension 
asymptotics of the distributions are essentially the same with detailed modifications in later sections.
In Sections~\ref{sec:rshigher} and \ref{sec:cs},
we study the joint distributions of arbitrary numbers of the eigenvectors of 
the real and complex symmetric random tensors, respectively. 
We obtain random matrix representations and large-dimension asymptotics of the joint distributions. 
In Section~\ref{sec:numerical} we perform some crosschecks of our results by carrying out 
Monte Carlo computations.
The final section is devoted to a summary and future prospects.

Lastly we would like to comment that 
the eigenvector distributions of symmetric random tensors can be considered to be
characteristic for tensors, since those of symmetric random matrices are trivial, as shown in \ref{app:trivial}. 

\section{Mean distribution of real eigenvectors of real symmetric random tensor}
\label{sec:rsmean}
In this section we study the mean distribution of the real eigenvectors of the real symmetric random tensor. 
The results themselves are not new, overlapping with \cite{pspin,randommat,Sasakura:2022iqd}, 
but we reproduce them to describe our procedure using some field theoretical methods in the 
simplest setting. The same procedure with detailed changes 
will be used to study the joint distributions in later sections. 

\subsection{Notations}
\label{sec:rsnotation}
Throughout this paper we take the convention that repeated indices are summed over, unless otherwise stated. 
We may call the usual (real and complex numbers) 
and Grassmann variables bosons and fermions, respectively, 
following standard naming in quantum field theories, 
though they have nothing to do with bosonic or fermionic particles.
Some variables may also be called fields, when they appear as variables in quantum field theoretical expressions,
though they have no spacetime dependence\footnote{Such fields are often called 
fields in zero dimensions.}.

Throughout Section~\ref{sec:rsmean} we particularly use the following conventions and short-hand notations:
\begin{itemize}

\item $T$ denotes a real symmetric order-$p$ dimension-$N$ tensor. Namely,  the components
$T_{a_1a_2\cdots a_p}\ (a_i=1,2,\cdots,N)$
satisfy $T_{a_1a_2\cdots a_p} \in \mathbb{R}$ and $T_{a_1a_2\cdots a_p}=T_{a_{\sigma[1]}a_{\sigma[2]}\cdots a_{\sigma[p]}}$ with $\sigma$ 
denoting arbitrary permutations of $\{ 1,2, \cdots, p\}$. The order of the tensor is restricted to be $p\geq 3$.  

\item The norm of a vector $x$ is defined by $|x|:=\sqrt{x_a x_a}$.

\item The delta function for a vector $x$ is defined by $\delta [x]:= \prod_{a=1}^N \delta [x_a]$, i.e., the product of the delta functions of each component.

\item Dot $\cdot$ between two vectors $x,y$ represents an inner product $x \cdot y:=x_a y_a$, and similarly for tensors, 
$X\cdot Y:=X_{a_1a_2\cdots a_p} Y_{a_1a_2\cdots a_p}$. When the two tensors are the same, we may write
$X\cdot X=:|X|^2$. We also use the dot for a partial contraction like 
$(X\cdot Y)_{a_1a_2\cdots a_p}:=X_{a_1a_2\cdots a_p b_1 b_2\cdots b_q} Y_{b_1b_2\cdots b_q}$
or
$(X\cdot Y)_{a_1a_2\cdots a_p}:= X_{b_1b_2\cdots b_q} Y_{b_1 b_2\cdots b_qa_1a_2\cdots a_p }$,
depending on the orders of the two tensors. 
 
\item Doubled square brackets are used to represent a symmetrization of a tensor, 
$\ss X\se_{a_1a_2\cdots a_q} 
:=\frac{1}{p!} \sum_{\sigma} X_{a_{\sigma[1]} a_{\sigma[2]} \cdots a_{\sigma[q]}}$, where 
the sum is over all the permutations of $\{1,2,\cdots,q \}$.

\item A tensor product of tensors (including vectors and matrices) 
is denoted by a simple sequence\footnote{This is usually denoted by $\otimes$, but using it would make 
equations too long and less visible in our paper. }. 
For instance, a tensor product of an order-$p$ tensor $X$ and 
an order-$q$ tensor $Y$ is written simply as $XY$, whose components are 
$(XY)_{a_1 a_2 \cdots a_{p+q}}:=X_{a_1\cdots a_p} Y_{a_{p+1}\cdots a_{p+q}}$. 
A tensor product of $n$ $X$'s may be written as $X^n \ (:=\overbrace{XX\cdots X}^n)$.

\item
The integration measure of a bosonic vector variable $x$ is defined by $dx:=\prod_{a=1}^N dx_a$,
unless otherwise stated. The integration region is over $\mathbb{R}^N$, unless otherwise stated.

\item The integration measure of an integration over $N$-dimensional Grassmann variables $\bpsi,\psi$ is 
defined so that $\int d\bpsi d\psi \, e^{\bpsi A \psi} = \det A$ for an arbitrary $N$-dimensional matrix $A$,
where $\bpsi A \psi:=\bpsi_a A_{ab} \psi_b$. For instance, see \cite{efetov} for more details.

\item $I_n$ denotes an $n$-dimensional identity matrix.

\item The surface volume of the unit sphere in an $n$-dimensional space is denoted by 
$S_{n-1}=2 \pi^{\frac{n}{2}}/ \Gamma\left[\frac{n}{2}\right]$,
where $\Gamma[\cdot]$ denotes the Gamma function. 

\end{itemize}

\subsection{Eigenvector distribution}
The real eigenpair equation of a real symmetric tensor $T$ is defined by \cite{Qi,lim,cart,qibook}
\[
T\cdot w^{p-1}=z \, w, \ |w|=1,
\label{eq:realepair}
\]
where $z\in \mathbb{R}$ and $w\in \mathbb{R}^N$.\footnote{In the usual index-wise notation, the eigenpair equation is 
given by $T_{a_1a_2\cdots a_p} w_{a_2} w_{a_3}\cdots w_{a_p}=z\, w_{a_1}$.}
$w$ and $z$ make an eigenpair of an eigenvector and an eigenvalue, respectively. 
Note that, for odd $N$, $z$ can always to be transformed to $z\geq 0$, by flipping the overall sign of $w$, 
if necessary. 
The eigenpair equation is invariant under 
\s[
T_{a_1a_2 \cdots a_p}&\rightarrow O_{a_1 a_1'}O_{a_2 a_2'}\cdots O_{a_p a_p'} T_{a'_1a'_2 \cdots a'_p},\\
w_a&\rightarrow O_{aa'} w_{a'},
\label{eq:rsortho}
\s]
where $O$ is an arbitrary real orthogonal matrix of dimension $N$. 

In our field theoretical treatment, the presence of the normalization condition $|w|=1$ 
of the eigenvector in \eq{eq:realepair} is not convenient.
Therefore we rather consider an {\it eigenvector} equation,
\[
 T \cdot v^{p-1}=v,
 \label{eq:egvec}
\]
where the eigenvalue is normalized instead of the eigenvector. 
Note that, unless $z=0$, \eq{eq:realepair} and \eq{eq:egvec} can be related.
By putting $v=|v| w$ into \eq{eq:egvec}, we obtain \eq{eq:realepair} 
with $z=|v|^{-p+2}$. Here $z$ becomes a positive value, but this is not a limitation. 
For odd $N$, an eigenvalue can be made $z> 0$, unless $z=0$, as explained above.
For even $N$, since the distribution of  $T$ we consider, which is the Gaussian distribution,
is symmetric under $T\rightarrow -T$, 
the distribution of $z$ is also symmetric under $z\rightarrow -z$. 
Therefore it is enough to compute the distribution of $z>0$.  
Moreover we can ignore $z=0$ probabilistically in practice, so we can obtain the distribution of 
the eigenvalue $z$ in \eq{eq:realepair} from the distribution of the eigenvector $v$ in \eq{eq:egvec}.

For a given $T$, the distribution of $v\ (\neq 0$, ignoring the trivial solution) from \eq{eq:egvec} is given by
\s[
\rho[v,T]&=\sum_{l=1}^{\# {\rm sol}[T]} \delta [v-v^{(l)}[T]] \\
&=| \det J[v,T] |\, \delta [f[v,T]], 
\label{eq:rsdefofrhovt}
\s]
where $v^{(l)}[T]\ (l=1,2,\cdots, \# {\rm sol}[T])$ are the non-zero solutions to \eq{eq:egvec} for $T$.
From the first line to the second, we have used that $v^{(l)}[T]$ are the solutions to $f[v,T]=0$ with a 
vector,
\[
f[v,T]:=v-T\cdot v^{p-1}.
\]
The factor $\left| \det J[v,T]\right|$ is the absolute value of the determinant of a Jacobian matrix,
\[
J[v,T]:= I_N- (p-1) \, T\cdot v^{p-2},
\label{eq:rsjacobian}
\]
which is from $J[v,T]_{aa'}=\frac{\partial f_{a'}}{\partial v_{a}}$. 
This absolute determinant factor is due to the transformation of the arguments of the delta functions
from the first to the second lines. The infinitesimal volume of $v$ associated to \eq{eq:rsdefofrhovt}
is $dv:=\prod_{a=1}^N dv_a$.

We consider the mean distribution of $\rho[v,T]$ under the Gaussian distribution of $T$, 
while the joint distributions will be studied later. More explicitly, it is defined by
\[
\rho[v]:= \left \langle \rho[v,T] \right\rangle_T := \frac{1}{A} \int dT\, e^{-\alpha\, |T|^2} \rho[v,T], 
\label{eq:realrhov}
\]
where $\int dT$ represents the integration of all the independent components\footnote{Note that 
components are related by $T_{a_1a_2\cdots a_p}=T_{a_{\sigma[1]}a_{\sigma[2]} \cdots a_{\sigma[p]}}$.
Therefore independents components are $T_{a_1a_2\cdots a_p}\ (a_1\leq a_2\leq \cdots \leq a_p$).} of $T$ over $\mathbb{R}$, $A:= \int dT\, e^{-\alpha\, |T|^2}$,
$|T|^2=T\cdot T=T_{a_1 a_2\cdots a_p}T_{a_1 a_2\cdots a_p}$, and $\alpha>0$. Note
that, since each independent component of $T$ may appear multiple times in $|T|^2$,
its variance expressed by \eq{eq:realrhov} depends on its multiplicity: 
$T_{\underbrace{\scriptstyle a_1\cdots a_1}_{n_1}\cdots \underbrace{\scriptstyle a_s\cdots a_s}_{n_s}}$ $(a_i<a_j$ for $i<j$) has
a variance $(\prod_{i=1}^s n_i !)/(2 \alpha p!)$.  

\subsection{Quantum field theoretical expression}
In this subsection we derive a field theoretical expression of \eq{eq:realrhov}. 
From a standard formula, $\delta [f ]=(2 \pi)^{-N} \int d\lambda \, e^{i \lambda \cdot f}$, where $\int d\lambda:=\int _{\mathbb{R}^N} \prod_{a=1}^N d\lambda_a$.   
As for $|\det J|$, we can consider 
\[
\left| \det J \right|&=\lim_{\epsilon\rightarrow +0} \frac{\det \left( J^2+\epsilon I_{N} \right) }{\sqrt{\det \left( J^2+\epsilon I_{N} \right)}}
=\lim_{\epsilon\rightarrow +0} \frac{\det \tilde J }{\sqrt{\det \tilde J}},
\label{eq:absdet}
\]
where $J^2$ is a matrix product of two $J$'s (differently from the notation in Section~\ref{sec:rsnotation}).
Here we have introduced a regularization parameter $\epsilon>0$ to avoid the possible singularity 
of the denominator, when $J$ contains zero eigenvalues, and 
\[
\tJ:=\left ( 
\begin{matrix}
\sqrt{\epsilon} I_N & i J \\
i J & \sqrt{\epsilon}  I_N
\end{matrix}
\right).
\] 

The denominator and the numerator on the righthand side of \eq{eq:absdet} 
can be expressed by the integrations over
$N$-dimensional bosonic variables $\phi_i\ (i=1,2)$ and fermionic ones $\bpsi_{i}, \psi_i\ (i=1,2)$, respectively.
The bosonic integration is over $\mathbb{R}^{2N}$, and the fermionic integration is defined in Section~\ref{sec:rsnotation}.
The expression can be made more compact by introducing 
superfields\footnote{When a field contains both bosonic and fermionic components, 
it is called a superfield in quantum field theories.}. 
Explicitly,
\[
|\det J|=& \lim_{\epsilon\rightarrow +0}
\frac{\det \tilde J }{\sqrt{\det \tilde J}}\cr
=& \lim_{\epsilon\rightarrow +0}\frac{1}{\pi^N} \int d\phi d\bpsi d\psi \exp \left[ 
\left( \begin{matrix}
\bpsi_1 & \bpsi_2 
\end{matrix}
\right)  \cdot \tilde J \cdot  
\left( \begin{matrix}
\psi_1 \\ \psi_2 
\end{matrix}
\right)
-
\left( \begin{matrix}
\phi_1 & \phi_2 
\end{matrix}
\right)  \cdot \tilde J \cdot  
\left( 
\begin{matrix}
\phi_1 \\ \phi_2 
\end{matrix}
\right)
\right] \cr 
=& \lim_{\epsilon\rightarrow +0} \frac{1}{\pi^N} \int d\Phi \exp \left[ 
i \G_{\kappa \kappa'}\, \Phi_\kappa \cdot J \cdot \Phi_{\kappa'}
 +\sqrt{\epsilon} \, \tG_{\kappa \kappa'} \Phi_\kappa \cdot \Phi_{\kappa'}
 \right],
 \label{eq:detexp}
 \]
where we have introduced a superfield $\Phi_\kappa$, whose index $\kappa$ denotes an index pair, $m=1,2,3$ 
and $i=1,2$, and defines 
$\Phi_{1i}:=\bpsi_i,\Phi_{2i}:=\psi_i,\Phi_{3i}:=\phi_i$. Here $d\Phi:=d\phi d\bpsi d\psi$, and  
\s[
&\G_{mi \, nj}:=K_{mn} \tK_{ij},  \\
&\tG_{mi\, nj}:= K_{mn}\delta_{ij},
\s] with 
\[
\K:=
\left( 
\begin{matrix}
0 & \frac{1}{2} & 0 \\
-\frac{1}{2} & 0 &0 \\
0 & 0& -1
\end{matrix}
\right)
,\ \ 
\tilde K:=
\left( 
\begin{matrix}
0 & 1 \\
1 & 0
\end{matrix}
\right)
.
\label{eq:kkt}
\]
Note that, while $\tilde K$ is symmetric, $K$ satisfies a graded symmetry $K_{mn}=(-1)^{\underline{m}\, \underline{n}} K_{nm}$, 
where $\underline{m}$ denotes the grade of the associated field,
namely, $\underline{1}=\underline{2}=1, \ \underline{3}=0$. Accordingly, 
$\G_{\kappa \kappa' }=(-1)^{\underline{\kappa}\, \underline{\kappa'}} \G_{\kappa' \kappa}$ and  
$\tG_{\kappa \kappa' }=(-1)^{\underline{\kappa}\, \underline{\kappa'}} \tG_{\kappa' \kappa}$, where $\underline \kappa$
is defined by $\underline \kappa=\underline{m}$ for $\kappa=mi$. Note also that 
$\Phi_\kappa \Phi_{\kappa'}=(-1)^{\underline{\kappa}\, \underline{\kappa'}} \Phi_{\kappa'} \Phi_{\kappa}$.

In \eq{eq:detexp} the bosonic and the fermionic parts are correlated:
\eq{eq:detexp} is symmetric under the following two supersymmetries $\dels_m\ (m=1,2)$ defined by 
\[
\dels_m \Phi_{ni}=\delta_{mn} \Phi_{3i},\ \dels_m \Phi_{3i}=K_{ml} \Phi_{li}, \ (m,n=1,2).
\label{eq:rssusy}
\]
\eq{eq:detexp} is also invariant under a discrete symmetry which interchanges $\Phi_{mi}$:
\[
\Phi_{m1} \leftrightarrow \Phi_{m2}.
\label{eq:rsintphi}
\]

By collecting the formulas above, the field theoretical expression of the eigenvector distribution is given by 
\[
\rho[v]=\lim_{\epsilon\rightarrow +0} \frac{1}{A (2\pi)^N \pi^N} \int dT d\lambda d\Phi \, e^{S},
\label{eq:rsrhoqft}
\]
where 
\[
S :=- \alpha |T|^2  +i \lambda\cdot  f[v,T] +i \G_{\kappa \kappa'}\, \Phi_\kappa \cdot J[v,T] \cdot \Phi_{\kappa'}
 +\sqrt{\epsilon} \, \tG_{\kappa \kappa'} \Phi_\kappa \cdot \Phi_{\kappa'}.
\label{eq:rss1}
\]

Let us lastly comment on the regularization.
The regularization parameter $\epsilon$ has been introduced for the 
quantum field theoretical computations to make the integration over the bosonic variables convergent. 
As a matter of fact, since $|\det J|$ is well-defined by itself, 
it does not have to be regularized for some other ways of computations. In fact,
the random matrix representations, which will be
discussed later, do not contain the regularization parameter. 
Also the Monte Carlo computations with explicitly solving the eigenvector equations performed in 
Section~\ref{sec:numerical} do not contain the determinant factor from the 
beginning\footnote{For instance in \eq{eq:rsdefofrhovt},
the determinant factor is needed to correct the non-linear $v$-dependence of the arguments of the delta functions.
On the other hand, if we just count the eigenvectors, as in the Monte Carlo simulations, 
there is no such a determinant factor, as in the first line of \eq{eq:rsdefofrhovt}.}.
 It is also proven in \ref{app:commute} that 
the distributions with no regularizations (equivalently, putting $\epsilon=0$) 
can be obtained by taking $\epsilon\rightarrow+0$ not only for finite $N$ but also 
after taking the asymptotic limit $N\rightarrow \infty$.
In addition, as will be shown in Section~\ref{sec:numerical}, 
the comparisons between the analytic large-$N$ asymptotic expressions and the 
Monte Carlo simulations numerically support the commutativity between
$\epsilon\rightarrow +0$ and $N\rightarrow \infty$, because the former is obtained by taking $\epsilon\rightarrow +0$
after $N\rightarrow \infty$, while the latter does not have the regularization.

\subsection{Integration over $T$ and $\lambda$}
In this subsection we perform the integration over $T$ and $\lambda$ in \eq{eq:rsrhoqft}. They appear at most quadratically in \eq{eq:rss1}, 
and therefore can explicitly be integrated over as Gaussian integrations. 

Let us pick up the terms containing $T$ in \eq{eq:rss1}:
\[
S_T:=-\alpha |T|^2 - i T\cdot \ss \lambda v^{p-1}\se -i (p-1) \G_{\kappa \kappa'}  T\cdot \ss \Phi_\kappa\Phi_{\kappa'} v^{p-2}\se.
\label{eq:rsst}
\]
Here we have used the short-hand notations listed in Section~\ref{sec:rsnotation}, 
$T\cdot \ss \lambda v^{p-1} \se=T_{a_1 a_2 \cdots a_p} 
\lambda_{a_1} v_{a_2} \cdots v_{a_p}$\footnote{The symmetrization of $\ss \lambda v^{p-1}\se$ is not needed here, because $T$ is symmetric.}
and for $T\cdot \ss \Phi_\kappa\Phi_{\kappa'} v^{p-2}\se$ as well.

It is convenient to decompose $\Phi_\kappa$ into the parallel and transverse parts against $v$:
\[
\Phi_{\kappa a} =\frac{v_a}{|v|} \Phi_{\kappa}^\parallel+ \Phi_{\kappa a}^\perp, 
\label{eq:rsdecompphi}
\]
where $\Phi_{\kappa}^\perp \cdot v=0$. Using this decomposition and the eigenvector equation \eq{eq:egvec}, 
we find that $T$ couples only with the transverse part:
\[
T\cdot \ss \Phi_\kappa\Phi_{\kappa'} v^{p-2}\se=\Phi_{\kappa}^\parallel \Phi_{\kappa'}^\parallel +T\cdot \ss \Phi_\kappa^\perp \Phi_{\kappa'}^\perp  v^{p-2}\se.
\label{eq:rsphidecouple}
\]
Here the reason why we can use \eq{eq:egvec} to derive \eq{eq:rsphidecouple} 
is that this term belongs to the determinant factor $J(v,T)$, which is 
multiplied with $\delta(f[v,T])$ imposing \eq{eq:egvec}, as in \eq{eq:rsdefofrhovt}.
On the other hand we cannot similarly decouple the parallel part of $\lambda$ from $T$, 
since the term $T\cdot \ss \lambda v^{p-1}\se$ in \eq{eq:rsst} belongs to $\delta(f[v,T])$ itself.
Putting \eq{eq:rsphidecouple} into \eq{eq:rsst}, the integration over $T$ results in  
\s[
\frac{1}{A}\int dT e^{S_T} =&\exp \left[ 
- i (p-1) \G_{\kappa \kappa'} \Phi_{\kappa}^\parallel \Phi_{\kappa'}^\parallel
-\frac{1}{4 \alpha}\left| \ss \lambda  v^{p-1} \se \right|^2 \right.  - \left. \frac{(p-1)^2}{4 \alpha} \left| \ss \G_{\kappa \kappa'}\Phi_\kappa^\perp \Phi_{\kappa'}^\perp  v^{p-2} \se \right|^2
 \right], 
 \label{eq:rsintT}
\s]
where we have used  $\ss \lambda  v^{p-1} \se \cdot \ss \Phi_\kappa^\perp \Phi_{\kappa'}^\perp  v^{p-2} \se=0$  
holding from $v\cdot \Phi_\kappa^\perp=0$.

Let us next perform the integration over $\lambda$. The terms containg $\lambda$ can
be collected from the second term of \eq{eq:rss1}\footnote{Note that the term $T\cdot \ss \lambda v^{p-1}\se$ in \eq{eq:rss1}
has already been included in $S_T$, and is therefore not included in $S_\lambda$.}
and from that of \eq{eq:rsintT}:
\[
S_\lambda&:=-\frac{1}{4 \alpha}\left|  \ss \lambda v^{p-1} \se \right|^2  + i \lambda\cdot v \cr
&= -\frac{|v|^{2(p-1)}}{ 4 \alpha} \left( \lambda^\parallel \right)^2 + i |v| \lambda^\parallel -\frac{|v|^{2(p-1)}}{4 p \alpha} 
\abs2{ \lambda^\perp},
\]
where we have introduced a similar decomposition, $\lambda_a=\frac{v_a}{|v|} \lambda^\parallel+\lambda^\perp_a$, into the parallel and the transverse
parts against $v$.  Then by integrating over $\lambda$ we obtain
\[
\int d\lambda\, e^{S_\lambda} =\left(\frac{4 \pi \alpha}{|v|^{2(p-1)}}\right)^\frac{N}{2} p^\frac{N-1}{2} \exp \left[-\frac{\alpha}{|v|^{2(p-2)}} \right].
\]

\subsection{Integration over $\Phi^\parallel_\kappa$} 
\label{sec:rsphipara}
The parallel part of the superfield $\Phi^\parallel_\kappa$ is contained in the third and the fourth terms of \eq{eq:rss1} and also 
exists in the first term of \eq{eq:rsintT}. By putting the decomposition \eq{eq:rsdecompphi} into \eq{eq:rss1} we obtain 
\[
S_\parallel:=& i \G_{\kappa \kappa'}\, \Phi_\kappa^\parallel \Phi_{\kappa'}^\parallel
 +\sqrt{\epsilon} \, \tG_{\kappa \kappa'} \Phi_\kappa^\parallel \Phi_{\kappa'}^\parallel 
- i (p-1) \G_{\kappa \kappa'} \Phi_{\kappa}^\parallel \Phi_{\kappa'}^\parallel  \cr
=& 
 i (2-p) \G_{\kappa \kappa'}\, \Phi_\kappa^\parallel \Phi_{\kappa'}^\parallel
 +\sqrt{\epsilon} \, \tG_{\kappa \kappa'} \Phi_\kappa^\parallel \Phi_{\kappa'}^\parallel. 
 \label{eq:rsspara}
\]
Note that \eq{eq:rsspara} has the same form as \eq{eq:detexp} with $J= 2-p$. Therefore,
\[
\lim_{\epsilon\rightarrow +0} \frac{1}{\pi} \int d\Phi_\parallel \, e^{S_\parallel}=p-2.
\]

The remaining variables are the transverse part of the superfield $\Phi^\perp_\kappa$.
The four-interaction term of $\Phi^\perp_\kappa$ in \eq{eq:rsintT} can be simplified as  
\[
\abs2{ \ss \G_{\kappa \kappa'} \Phi_\kappa^\perp \Phi_{\kappa'}^\perp  v^{p-2} \se }
=\frac{2 |v|^{2(p-2)}}{p(p-1)} \abs2{\ss \G_{\kappa \kappa'} \Phi_\kappa^\perp \Phi_{\kappa'}^\perp\se} 
\label{eq:rsintsimp}
\]
by performing the symmetrization and using $v \cdot \Phi_\kappa^\perp=0$.
In this derivation it has been useful to notice the symmetry,  
$\G_{\kappa \kappa'} \Phi^\perp_{\kappa a}  \Phi^\perp_{\kappa' b} =\G_{\kappa \kappa'}  \Phi^\perp_{\kappa b} \Phi^\perp_{\kappa' a}$, which holds 
because the signatures of the graded symmetries of $\G_{\kappa \kappa'}$ and $\Phi_\kappa$ 
cancel with each other (See below \eq{eq:kkt}). 
Then, by putting the decomposition \eq{eq:rsdecompphi} into \eq{eq:rss1}, including the interaction term \eq{eq:rsintsimp}, and 
putting the results of the integrations over $T,\lambda,\Phi^\parallel_\kappa$ into \eq{eq:rsrhoqft}, we obtain
\[
\rho[v] =  (p-2)\, p^\frac{N-1}{2} \left( \frac{\alpha}{\pi |v|^{2(p-2)}}\right)^\frac{N}{2} |v|^{-N}  
\exp\left[ -\frac{\alpha}{|v|^{2 (p-2)}} \right] Z_\perp,
\]
where 
\[
Z_\perp:=\lim_{\epsilon\rightarrow +0} \frac{1}{\pi^{N-1}} \int d\Phi^\perp \, e^{S_\perp}
\label{eq:rszperp}
\]
with
\[
S_\perp:=i \G_{\kappa \kappa'}\, \Phi_\kappa^\perp \cdot \Phi_{\kappa'}^\perp
 +\sqrt{\epsilon} \, \tG_{\kappa \kappa'} \Phi_\kappa^\perp  \cdot \Phi_{\kappa'}^\perp -
 \frac{ (p-1) |v|^{2(p-2)}}{2 p \alpha} \abs2{\ss \G_{\kappa \kappa'}\Phi_\kappa^\perp \Phi_{\kappa'}^\perp\se}
\label{eq:rssperp}
\]
Here $Z_\perp$  is normalized for the non-interacting case $|v|=0$:
\[
\left.  Z_\perp \right | _{|v|=0}=1.
\label{eq:rsnormz}
\]

Since $\rho [v]$ depends only on $|v|$, it is more appropriate to integrate $v$ over the angular directions. 
Then we obtain the size distribution of the eigenvectors as
\[
\bar \rho \left[|v|\right]:=&\int_{\mathbb{R}^N} dx\, \delta\left[ |v|-|x| \right] \, \rho(x) \cr
=&2 (p-2)\, p^\frac{N-1}{2}  \Gamma[N/2]^{-1} \left( \frac{\alpha}{|v|^{2 (p-2)}} \right)^\frac{N}{2} |v|^{-1} 
\exp\left[ -\frac{\alpha}{|v|^{2 (p-2)}} \right] Z_\perp,
\label{eq:rsrhobar}
\]
where we have used the formula of the volume of $N-1$-dimensional unit sphere $S_{N-1}$
in Section~\ref{sec:rsnotation}. The associated infinitesimal volume is $d|v|$.

Since the eigenvalue $z$ has the relation, $z=|v|^{-p+2}$ (see below \eq{eq:egvec}), the eigenvalue distribution is given 
by\footnote{From $\rho_{\rm eigenvalue}[z]=\bar \rho [|v|] \left| \frac{d|v|}{dz} \right|$.}
\[
\rho_{\rm eigenvalue}[z]=2 p^{\frac{N-2}{2}} \Gamma[N/2]^{-1} (\alpha z^2)^\frac{N}{2} z^{-1} \exp[-\alpha z^2] 
\left. Z_\perp \right|_{|v|=z^{-1/(p-2)}},
\]
whose associated infinitesimal volume is $dz$.

\subsection{Random matrix representation of $Z_{\perp}$}
\label{sec:rsranmat}
The interaction term of $S_\perp$ in \eq{eq:rssperp} has the form $-\gc H_{ab} H_{ab}$ with a symmetric matrix $H$, 
where $H:=\G_{\kappa \kappa'} \Phi^\perp_{\kappa} \Phi^\perp_{\kappa'}$ and 
\[
\gc:=\frac{ (p-1) |v|^{2(p-2)}}{2 p \alpha}.
\label{eq:rsdefg}
\]
We introduce an auxiliary matrix variable $M$ to rewrite the interaction term in linear in $H_{ab}$ (quadratic in $\Phi^\perp_\kappa$)\footnote{This procedure is called Hubbard-Stratonovich transformation in
the quantum field theory.}:
\[
\exp[-\gc H_{ab} H_{ab} ]=\frac{1}{Z_{\rm RM}^0}\int_{V_M} dM \exp \left[ -M_{ab} M_{ab} - 2  i \sqrt{\gc} M_{ab} H_{ab} \right],
\]
where the integration is over the $(N-1)$-dimensional symmetric real matrix $M$ in the transverse subspace 
against $v$, and $Z_{\rm RM}^0 := \int_{V_M} dM \exp \left[ -M_{ab} M_{ab} \right]$. 
Then, by applying this transformation to the interaction term of 
\eq{eq:rssperp}\footnote{In such a case that $H_{ab}$ contains fermionic variables as well, 
we need to see whether there are relevant contributions from integration boundaries of bosonic parts of 
auxiliary variables. 
In the current case, however, there are no boundary contributions, sincee
the integrand damps quickly toward infinites in $\sim \exp(-M^2)$.
For instance, see \cite{efetov} for more details.},  the action of $\Phi_\kappa^\perp$ is obtained as
\[
S_{\perp M}:=
i \G_{\kappa \kappa'}\, \Phi_\kappa^\perp \cdot (I- 2  \sqrt{\gc} M)\cdot \Phi_{\kappa'}^\perp
 +\sqrt{\epsilon} \, \tG_{\kappa \kappa'} \Phi_\kappa^\perp  \cdot \Phi_{\kappa'}^\perp.
 \label{eq:rssphiperp}
\]
Since this has the form of \eq{eq:detexp} with $J=I- 2  \sqrt{\gc} M$,
the integration over $\Phi_\kappa^\perp$ results in a random matrix expression, 
\[
Z_\perp = \frac{1}{Z_{\rm RM}^0}\int_{V_M} dM \left| 
\det \left(I- 2  \sqrt{\gc} M \right) \right| \exp \left[  -M_{ab} M_{ab} \right].
\label{eq:rszmat}
\]

The random matrix model \eq{eq:rszmat} can be 
expressed by an integral over the matrix eigenvalues $z_i\ (i=1,2,\cdots, N-1)$ of $M$. 
More explicitly, it is expressed by using the correlation functions $P^{(n)}(\cdots)$
of the matrix eigenvalues:
\[
Z_\perp
=\frac{(2\gc)^\frac{N-1}{2} e^{z_{N}^2/2} \int_0^\infty dz_1 \cdots dz_{N-1}\, P^{(N)}(z_1,\cdots,z_{N})}
{\int_0^\infty dz_1 \cdots dz_{N-1}\, P^{(N-1)}(z_1,\cdots,z_{N-1})}, 
\label{eq:rszperpexact}
\] 
where $z_{N}=-1/\sqrt{2\gc}$, and 
\s[
P^{(n)}(z_1,\cdots,z_{n}) :=e^{ - \sum_{i=1}^n z_i^2/2} \prod_{i=1,j=2 \atop i<j}^n  \left| z_i-z_j \right|. 
\s]
It is known that these integrals can explicitly be performed by using skew orthogonal 
polynomials. For self-containedness of this paper, 
the explicit functional forms are given in \ref{app:rsexplicitI}.

\subsection{Large-$N$ limit}
\label{sec:rslargen}
A large-$N$ limit can be taken for $\bar \rho\left[|v|\right]$ with a scaling $v=N^{-1/(2(p-2))} \tv$, where $\tv$ is kept 
$O(1)$ in the limit.
While it is straightforward to take the limit for the overall factor of \eq{eq:rsrhobar}, it is a non-trivial task to take it for 
the quantum field theoretical part defined by \eq{eq:rszperp} and \eq{eq:rssperp}. 
A physicist-friendly method for taking the limit is to use the Schwinger-Dyson equations.  
In the lowest order of the method, we solve some consistency equations of the two-field correlation 
functions, 
assuming that these correlation functions give the leading-order contributions.
(For instance, see an appendix in \cite{Sasakura:2022iqd} for a quick account of the method.)

Let us assume that the symmetries \eq{eq:rssusy} and \eq{eq:rsintphi} are not violated 
in the large $N$-limit. Note also that the orthogonal symmetry (a consequence of \eq{eq:rsortho}) 
holds in the transverse subspace. 
Then, as shown in \ref{app:susytwo}, the two-field correlation functions can uniquely be determined to have the form, 
\s[
\langle \pPhi_{mia} \pPhi_{njb} \rangle = K^{-1}_{mn} Q_{ij} I^\perp_{ab},
\label{eq:rsassq}
\s] 
where $I^{\perp}$ is the identity matrix in the transverse subspace against $v$, and 
\[
Q:=
\left(
\begin{matrix}
Q_1 & Q_2 \\
Q_2 & Q_1
\end{matrix}
\right),
\label{eq:theformofQ}
\]
which will be determined by solving the equations below. 
Using \eq{eq:rsassq} to approximate the expectation value $\langle S_\perp \rangle $ of \eq{eq:rssperp}, 
the effective action in the leading order of $N$ can be computed as
\[
S_{\rm eff}=&\left\langle S_{\perp} \right\rangle - \frac{1}{2} \log \sdet \langle \pPhi_{mia} \pPhi_{njb} \rangle, \cr
= & N \left(- 2 i\,  Q_2-2 \sqrt{\epsilon} \, Q_1 +2 \tgc \left( Q_1^2+Q_2^2 \right) - \frac{1}{2} \log \left( Q_1^2-Q_2^2 \right) 
+\hbox{constant}\right)
+o(N),
\label{eq:rsseff}
\] 
where $\sdet \langle \pPhi_{mia} \pPhi_{njb} \rangle$ 
denotes the superdeterminant\footnote{In the current case, since the fermion-boson correlation functions vanish, 
\[
\sdet \left( 
\begin{matrix}
A & 0 \\
0 & B
\end{matrix}
\right)=\frac{\det A}{\det B},
\] 
where $A$ and $B$ are the fermion-fermion and boson-boson correlation functions, respectively.}
of a matrix $\langle \pPhi_{mia} \pPhi_{njb} \rangle$ with $mia$ and $njb$ being respectively 
the column and row indices, and $\tgc:= \gc N$, which is kept $O(1)$ 
in the large-$N$ limit (consistent with $v=N^{-1/(2(p-2))} \tv$ above). 
Here the constant term is unimportant, because it will be taken later 
to be consistent with the normalization \eq{eq:rsnormz}.     
Note also that we have ignored the difference of dimensions between $N$ and $N-1$ due to the large $N$ limit. 
Some details of the computation of $S_{\rm eff}$ are given in \ref{app:SD}.

$Q$ is determined by solving the stationary conditions $\frac{\partial S_{\rm eff}}{\partial Q_1}=\frac{\partial S_{\rm eff}}{\partial Q_2}=0$. This 
gives a system of cubic equations, which can be solved analytically. 
There are four independent solutions. Among them,  acceptable is the one which is consistent 
with the two-field correlation functions at the non-interacting case $\gamma=0$.
This requires $Q_1=0, Q_2=i/2$ for $\tgc=0, \epsilon\rightarrow +0$. 
This condition uniquely selects a solution among the four. By taking $\epsilon \rightarrow +0$ 
of the unique solution, we obtain
\[
&Q_1=0,\ Q_2=\frac{i(1-\sqrt{1-4 \tgc}) }{4 \tgc} \hbox{   for } \tgc\leq \tgc_c,  \\
&Q_1=-\frac{\sqrt{4\tgc-1}}{4 \tgc},\ Q_2=\frac{i}{4 \tgc}  \hbox{   for } \tgc\geq \tgc_c,
\]
with $\tgc_c=1/4$. $\tgc_c$ is a phase transition point separating the weak and strong coupling regimes of $Z_\perp$.  
This point corresponds to $E_\infty$ of \cite{randommat}, 
which separates the parameter regions where the critical points of the energy landscape are dominated
by stable ones or by unstable ones, respectively, in the spherical $p$-spin spin glass model \cite{pspin,pedestrians}. 
By putting these solutions to \eq{eq:rsseff}, we obtain in the large-$N$ limit 
\[
Z_\perp&=\exp\left[ S_{\rm eff}-\left. S_{\rm eff} \right|_{\tgc=0}+o(N)\right] \cr
&= \left\{ 
\begin{matrix}
\exp \left[ 
N\left( -\frac{1}{2}  +\frac{1-\sqrt{1-4 \tgc}}{4 \tgc} -\log \left[ \frac{1-\sqrt{1-4\tgc}}{2 \tgc} \right] \right)+o(N) \right] & \hbox{for } \tgc\leq \tgc_c \\
\exp \left[ N\left( -\frac{1}{2} +\frac{1}{4 \tgc} +\frac{1}{2} \log [\tgc] \right)+o(N) \right] & \hbox{for } \tgc\geq \tgc_c
\end{matrix}
\right.,
\label{eq:zperplargen}
\]
where we have tuned the normalization to be consistent with \eq{eq:rsnormz} in order $N$ 
by subtracting the value of $S_{\rm eff}$ at $\tgc=0$.

By combining with the large-$N$ asymptotics of the overall coefficient of \eq{eq:rsrhobar}, we obtain
\[
\rho\left[ |v| \right]=\exp \left[ N\, {\rm Re}\left[ h_p[\tgc/\tgc_c]\right]+ o(N)\right], 
\label{eq:rsrhoabs}
\]
where $\tgc/\tgc_c=2(p-1) |\tilde v|^{2(p-2)}/(p \alpha)=2(p-1)N |v|^{2(p-2)}/(p \alpha)$, 
${\rm Re}[\cdot]$ is to take the real part, and $h_p[\cdot]$ is a universal function obtained in \cite{Delporte:2025tjp} 
across various tensor eigenvalue distributions:
\[
h_p[x]=\frac{1}{2} \log[p-1] +\frac{1}{x} \left( -1+\frac{2}{p} -\sqrt{1-x} \right) 
+\frac{1}{2}\log[x] -\log\left[1-\sqrt{1-x}\right].
\]

Lastly let us comment on the order of the two limits taken in this section. 
The large-$N$ asymptotic limit is taken first to compute the effective action,
and the $\epsilon\rightarrow +0$ limit is taken next after finding the unique solution. This order 
is essential to obtain a unique solution over the whole region of $\tilde \gamma$,
satisfying the boundary condition at $\tilde \gamma=0$. 
If we took the two limits in the reverse order, the singularity at $\tilde \gamma=\tilde \gamma_c$ would
block extending uniquely the solution in $\tilde \gamma<\tilde \gamma_c$ to the region 
$\tilde \gamma>\tilde \gamma_c$.
Since the genuine definition of the distribution does not contain the regularization, which is 
equivalent to taking $\epsilon\rightarrow +0$ first, we need a justification for the order of the two limits 
we have taken in this section.
This is given in \ref{app:commute}. The numerical crosschecks in Section~\ref{sec:numerical} also give 
an independent justification for the commutativity of the two limits. 

\section{Joint distributions of real eigenvectors of real symmetric random tensor}
\label{sec:rshigher}
In this section we extend the computation of Section~\ref{sec:rsmean} to the joint distributions of 
the real eigenvectors of the real symmetric random tensor.  

\subsection{Notations}
\label{sec:rshighernot}
We inherit the conventions and the short-hand notations listed in Section~\ref{sec:rsnotation}.
We consider sets of eigenvectors $v_l\ (l=1,2,\cdots,L)$, where $L\leq N$. 
Since we consider multiple eigenvectors, we introduce the following additional notations:
\begin{itemize}
\item
The vector subspace $V_\parallel$ is the vector space spanned by the eigenvectors $v_l\ (l=1,2,\cdots,L)$. 
We assume $v_l$ are linearly independent. 
This is a reasonable assumption for $L \leq N$ for an arbitrarily given $T$.     
$V_\perp$ is the vector subspace transverse to $V_\parallel$: $v\cdot v'=0$ for ${}^\forall v\in V_\parallel, {}^\forall v'\in V_\perp$. 
We denote an arbitrary set of linearly independent vectors spanning $V_\perp$ as $v^\perp_m\ (m=1,2,\cdots,N-L)$. 
$V=V_\parallel \oplus V_\perp$.
\item 
Metrics
\s[
&g_{mn}:=v_m \cdot v_n, \\
&g^\perp_{mn}:=v_m^\perp \cdot v_n^\perp, \\
&\gq{q}{m}{n}:= v_m^{q} \cdot v_n^{q}=(g_{mn})^{q}, \\
&\gt{mn}{m'n'}:=\ss v_m v_n^{p-1}\se \cdot \ss v_{m'} v_{n'}^{p-1} \se=\frac{1}{p} \left( 
g_{mm'} \gq{p-1}{n}{n'}+(p-1) g_{mn'} g_{nm'} \gq{p-2}{n}{n'} \right). \\
&\gtp{mn}{m'n'}:=\ss v^\perp_m v_n^{p-1}\se \cdot \ss v^\perp_{m'} v_{n'}^{p-1}\se=\frac{1}{p} g^\perp_{mm'} \gq{p-1}{n}{n'}.
\s]
\item
Projectors on symmetric order-$p$ tensors
\s[
&\Ipara:=\ss v_mv_n^{p-1} \se \gtinv{mn}{m'n'} \ss v_{m'} v_{n'} ^{p-1}\se, \\
&\Iperp:=I-\Ipara, \\
&\tIpara:=\ss v^\perp_mv_n^{p-1} \se \gtpinv{mn}{m'n'} \ss v^\perp_{m'} v_{n'} ^{p-1}\se, \\
&\tIperp:=I-\Ipara-\tIpara.
\s] 
$\gtinv{}{}$ and $\gtpinv{}{}$ denote the inverse matrices of $\gt{}{}$ and $\gtp{}{}$, respectively, 
where they are regarded as matrices with row index $mn$ and column index $m'n'$.  
$I$ denotes an identity projector, and the projectors operate, for example, as $I\cdot T=T$ and 
$\Ipara \cdot T=\ss v_mv_n^{p-1} \se \gtinv{mn}{m'n'} \ss v_{m'} v_{n'} ^{p-1}\se\cdot T$
on an arbitrary order-$p$ symmetric tensor $T$. Hence $\Ipara$ and $\tIpara$ are the projections on 
the symmetric tensor subspaces spanned by $\ss v_mv_n^{p-1} \se\ (m,n=1,2,\cdots,L)$ and $\ss v^\perp_m v_n^{p-1} \se\ (m=1,2,\cdots,N-L, n=1,2,\cdots,L)$, respectively. 

\end{itemize}

\subsection{Quantum field theoretical expression}
\label{sec:hmqft}
The joint distributions of a set of eigenvectors $\{v_1,v_2,\cdots, v_L \}$ are defined by
\[
\rho_L[\{v\}]:=\left\langle \prod_{l=1}^L \rho[v_l,T] \right\rangle_T,
\label{eq:rsrhoLdef}
\]
where $\rho[\cdot,\cdot]$ is defined in \eq{eq:rsdefofrhovt}, $\{v\}$ is a short-hand notation for a set 
of eigenvectors, and $\langle \cdot \rangle_T$ is to take the mean over 
the Gaussian distribution of $T$ defined in \eq{eq:realrhov}. This expression is not well-defined, when 
there exist coincident eigenvectors, because of products of the same delta-functions.
This possibility is removed by the assumption of the linear independence of $v_l$, 
as mentioned in Section~\ref{sec:rshighernot}.
By proceeding in a similar manner as before, we obtain a quantum field theoretical expression
by simply multiplying the variables:
\[
\rho_L[\{v\}]
=\lim_{\epsilon\rightarrow +0} \frac{1}{A (2\pi)^{N L}\pi^{N L}} \int dT d\lambda d\Phi \, e^{S_L}
\label{eq:rsrhoqftL}
\]
with
\[
S_L: =- \alpha\abs2{T}  +i \lambda_l \cdot  f[v_l,T] +i \G_{\kappa \kappa'}\, \Phi_{l \kappa} \cdot J[v_l,T] \cdot \Phi_{l \kappa'}
 +\sqrt{\epsilon} \, \tG_{\kappa \kappa'} \Phi_{l \kappa}  \cdot \Phi_{l\kappa'},
\label{eq:rssmulti}
\]
where we have introduced multiple real vector variables $\lambda_l$ 
as well as superfields $\Phi_{l\kappa}\ (l=1,2,\cdots,L)$.
The integration measure follows the previous case by simple multiplications.

The system \eq{eq:rsrhoqftL} with \eq{eq:rssmulti} is invariant under the supersymmetry \eq{eq:rssusy} and the discrete symmetry
 \eq{eq:rsintphi} for each sector of $\Phi_{l}$. More explicitly they are given by
\[
\dels_{lm} \Phi_{l'ni}=\delta_{ll'} \delta_{mn} \Phi_{l3i},\ \dels_{lm} \Phi_{l'3i}=\delta_{ll'} K_{mp} \Phi_{lpi}, \ (l,l'=1,2,\cdots,L,\ m,n=1,2),
\label{eq:rssusyL}
\]
and
\[
\Phi_{lm1} \leftrightarrow \Phi_{lm2},
\label{eq:rsintphiL}
\]
respectively.

\subsection{Integration over $T$ and $\lambda$}
\label{sec:rsinttlamL}
The process of computations is similar to that of the mean distribution. 
In this subsection we explicitly integrate over $T$ and $\lambda$ in \eq{eq:rsrhoqftL}. 
The terms containing $T$ in \eq{eq:rssmulti} are given by 
\[
S_{LT}:=-\alpha \abs2{T} -i \,T\cdot  \ss \lambda_l  v_l^{p-1} + (p-1) \G_{\kappa \kappa'} \Phi_{l \kappa} \Phi_{l \kappa'} v_l^{p-2}
\se.
\label{eq:slt}
\]

Let us first show that 
$T$ decouples from a parallel part of $\Phi_{l\kappa}$, as in the previous case.  Let us first note 
\[
T\cdot \ss v_m v_n^{p-1} \se=T_{a_1a_2\cdots a_p} v_{ma_1} v_{na_2}\cdots v_{na_p}=v_m\cdot v_n=g_{mn},
\]
where we have used the eigenvector equation \eq{eq:egvec} and the metric $g$ defined 
in Section~\ref{sec:rshighernot}. 
Therefore,
\[
&T\cdot  \ss \G_{\kappa \kappa'} \Phi_{l \kappa} \Phi_{l \kappa'} v_l^{p-2}\se=
T\cdot \left( \Ipara +\Iperp \right) \cdot \ss \G_{\kappa \kappa'} \Phi_{l \kappa} \Phi_{l \kappa'} v_l^{p-2}\se \cr
&\hspace{1cm}
=Y \cdot \ss \G_{\kappa \kappa'} \Phi_{l \kappa} \Phi_{l \kappa'} v_l^{p-2}\se+
T\cdot \Iperp \cdot \ss \G_{\kappa \kappa'} \Phi_{l \kappa} \Phi_{l \kappa'} v_l^{p-2}\se,
\]
where we have used the projectors in Section~\ref{sec:rshighernot}, and 
\[
Y:= g_{mn}\gtinv{mn}{m'n'} \ss v_{m'} v_{n'}^{p-1} \se.
\label{eq:rsY}
\] 
We see that $T$ couples only with the tensor subspace
projected by $\Iperp$. 
Putting this expression into \eq{eq:slt} and performing the integration over $T$, we obtain
\s[
\frac{1}{A}\int dT\,  e^{S_{LT}}=\exp \bigg[&-i (p-1)\, Y \cdot \ss \G_{\kappa \kappa'} \Phi_{l \kappa} \Phi_{l \kappa'} v_l^{p-2}\se \\
&\ \ \ -\frac{1}{4\alpha} \abs2{\ss \lambda_l v_l^{p-1}\se +(p-1) \Iperp\cdot 
\ss G_{\kappa \kappa'} \Phi_{l \kappa} \Phi_{l \kappa'} v_l^{p-2}\se } \bigg].
\s]

Let us next perform the integration over $\lambda$. The terms containing $\lambda$ are
\s[
S_{L\lambda}&:= i \lambda_l \cdot v_l  -\frac{1}{4\alpha} \abs2{\ss \lambda_l v_l^{p-1} \se +(p-1) \Iperp\cdot 
 \ss G_{\kappa \kappa'}\Phi_{l \kappa} \Phi_{l \kappa'} v_l^{p-2} \se} \\
 &= i \lambda_l^\parallel \cdot v_l  -\frac{1}{4\alpha} \abs2{\ss \lambda_l^\parallel v_l^{p-1} \se +
 \ss \lambda_l^\perp v_l^{p-1} \se +(p-1)(\tIpara+\tIperp)\cdot 
 \ss G_{\kappa \kappa'}\Phi_{l \kappa} \Phi_{l \kappa'} v_l^{p-2} \se},
\label{eq:sllam}
\s]
where we have made the splitting $\lambda_l=\lambda_l^\parallel+\lambda^\perp_l$, where 
$\lambda_l^\parallel \in V_\parallel$ and $\lambda_l^\perp \in V_\perp$, and 
have used $\Iperp=\tIpara+\tIperp$.
Now, by performing a shift of $\lambda_l^\perp$, one can remove 
$\tIpara \cdot \ss  G_{\kappa \kappa'} \Phi_{l \kappa} \Phi_{l \kappa'} v_l^{p-2} \se$,
since this is a linear combination of $\ss v_m^\perp v_n^{p-1}\se.$
Then, since the remaining terms are orthogonal to each other, we find 
\s[
S_{L\lambda}=&i \lambda_l^\parallel \cdot v_l-\frac{1}{4 \alpha} \abs2{ \ss  \lambda_l^\parallel  v_l^{p-1} \se}-\frac{1}{4 \alpha}
\abs2{\ss \lambda^\perp_l v_l^{p-1} \se} \\
&\ \ \ -\frac{(p-1)^2}{4 \alpha}  \ss \G_{\kappa \kappa'} \Phi_{l \kappa} \Phi_{l \kappa'} v_l^{p-2} \se \cdot \tIperp \cdot
\ss G_{\kappa'' \kappa'''} \Phi_{l' \kappa''} \Phi_{l' \kappa'''} v_{l'}^{p-2} \se.
\s]

The explicit integrations over $\lambda_l$ can be straightforwardly performed, for instance, by expanding $\lambda_l^\parallel=
\lambda^\parallel_{lm}v_m$. Then we obtain
\s[
&\rho_L[\{v\}]=\\
&\ \ \left(\frac{\alpha}{\pi}\right)^{\frac{NL}{2}} p^\frac{L(N-L)}{2}  \left( \det g \right)^\frac{L}{2} 
\left(\det \gt{}{} \right)^{-\frac{1}{2}} \left( \det \gq{p-1}{}{} \right)^{-\frac{N-L}{2}}
\exp\left[ 
-\alpha\, g\cdot  \gtinv{}{} \cdot g 
\right]
Z_L,
\label{eq:rscoeffL}
\s]
where
\[
Z_{L}:=\lim_{\epsilon\rightarrow +0} \frac{1}{\pi^{NL}}\int d\Phi\, e^{S_{L\Phi}}
\]
with
\s[
S_{L\Phi}
:=&i \G_{\kappa \kappa'} \Phi_{l \kappa}\cdot \Phi_{l \kappa'} +\sqrt{\epsilon} \tG_{\kappa \kappa'} \Phi_{l \kappa}\cdot \Phi_{l \kappa'}  
-i (p-1) \, Y\cdot \ss \G_{\kappa \kappa'} \Phi_{l \kappa} \Phi_{l \kappa'} v_l^{p-2} \se \\
&\ 
-\frac{(p-1)^2}{4 \alpha}  \ss \G_{\kappa \kappa'} \Phi_{l \kappa} \Phi_{l \kappa'} v_l^{p-2} \se \cdot \tIperp \cdot
\ss G_{\kappa'' \kappa'''} \Phi_{l' \kappa''} \Phi_{l' \kappa'''} v_{l'}^{p-2} \se.
\label{eq:rssLp}
\s]

Since the vector space indices are all contracted in \eq{eq:rscoeffL} and \eq{eq:rssLp},  $\rho_L[\{v\}]$ is symmetric
under the orthogonal transformation in $V$.
To explore the real properties of $\rho_L[\{v\}]$, it is appropriate to remove the redundancy, 
extending the rotationally symmetric expression \eq{eq:rsrhobar} for the mean distribution. 
As shown in \ref{app:volume}, a convenient parametrization is given by 
\[
v_{la}= 0 \hbox{ for } a > l, \ v_{ll}>0,
\label{eq:volgauge}
\]
where $v_{la}\ (l<a)$ can take any real numbers, $v_{ll}$ any positive numbers, 
and the associated infinitesimal volume is given by
\[
dv':=\prod_{l=1}^L \left( S_{N-l} \, v_{ll}^{N-l} \prod_{a=1}^l dv_{la}\right).
\label{eq:volfac}
\]
With this parametrization $\{ v\}'$ the eigenvector distribution is given by
\s[
\bar \rho_L[\{v\}']&=\left(\prod_{l=1}^L  S_{N-l} \, v_{ll}^{N-l} \right) \rho_L[\{v\}].
\label{eq:rhobarL}
\s]

As in Section~\ref{sec:rsphipara}, one can decouple the component of $\Phi_{l\kappa}$ parallel to $v_l$ 
from the interaction term in \eq{eq:rssLp} 
and can explicitly be integrated over. However, this does not largely help the subsequent 
computations, since the other components of $\Phi_{l\kappa}$ 
belonging to $V_\parallel$ still exist in the interaction term. 
Therefore we will not perform this process for this case. Interested readers can see \ref{app:rsphipara} for this process. 

\subsection{Random matrix representation of $Z_L$}
 As in Section~\ref{sec:rsranmat}, we can introduce auxiliary variables to transform the four-interaction 
 term in \eq{eq:rssLp} to a
quadratic form in $\Phi_{l\kappa}$. A difference from the previous case is that,
because of the intricacy of the interaction term, the choice of auxiliary variables is ambiguous and less explicit.
However, the following formulas give an enough basis 
for the numerical study in Section~\ref{sec:numerical}. 

Let us consider 
\[
Z_{\rm {\it L\,}RM}:=\int_{V_M} dM \exp \left[
-\ss M_l v_l^{p-2} +i H_l v_l^{p-2} \se \cdot \tIperp \cdot \ss  M_{l'} v_{l'}^{p-2} +i H_{l'} v_{l'}^{p-2} \se
\right],
\label{eq:zm}
\]
where $M_{l}\ (l=1,2,\cdots,L)$ are real symmetric matrices of dimension $N$ and
 \[
 H_l := \frac{p-1}{2 \sqrt{\alpha}} \G_{\kappa \kappa'} \Phi_{l \kappa} \Phi_{l \kappa'}.
 \]  
Here the integration region of $M$ is taken to be a vector space $V_M$, and one requirement is that $V_M$ should not contain 
an element $M$ satisfying $\tIperp \cdot \ss  M_{l} v_{l}^{p-2} \se=0$ to avoid divergence of the integration.
Therefore we take a quotient vector space,
\[
V_M :\cong V_{RS}/K_{\tIperp\cdot  v^{p-2} },
\label{eq:defofvm}
\]
where $V_{RS}$ is the vector space of $L$  real symmetric dimension-$N$ matrices,
and $K_{\tIperp\cdot  v^{p-2} }$ 
is a kernel, 
\[
K_{\tIperp\cdot  v^{p-2} }:=\{(M_1,M_2,\cdots,M_L ) \,
|\,  \tIperp\cdot \ss  M_{l} v_{l}^{p-2} \se=0, (M_1,M_2,\cdots,M_L )\in V_{RS} \}.
\label{eq:rskernel}
\]

An explicit 
choice of such a $V_M$ is to take the vector subspace transverse to the kernel $K_{\tIperp\cdot  v^{p-2} }$, but 
other choices are also allowed.
Note that \eq{eq:zm} has the same value as $Z_{\rm {\it L\,}RM}^0:=Z_{\rm {\it L\,}RM}|_{H_l=0}$.
This can be shown by regarding the integration as a holomorphic complex integration and 
shifting the integration variables $M_l$ by imaginary numbers to remove 
$\tIperp \cdot \ss H_l v_l^{p-2} \se$\footnote{Note that it is not needed to exactly cancel $H_l$ 
by performing a shift of $M_l$. This may not be possible in general, depending on $V_M$ and $H_l$. 
What should be removed is the projected quantity $i \tIperp \cdot
\ss H_lv_l^{p-2} \se$ by $\tIperp \cdot \ss M_lv_l^{p-2} \se$.
This is always possible, if one chooses $V_M$ as in \eq{eq:defofvm}, since the latter spans all the possibilities.
Shifting by imaginary numbers should not also be a problem, 
analogously to $\int dx \exp(-x^2)=\int dx \exp[-(x-i a )^2]$.
One can also take into account that $H_l$ contains fermionic variables (See a footnote in Section~\ref{sec:rsranmat}). }. 
By multiplying $1=Z_{\rm {\it L\,}RM}/Z_{\rm {\it L\,}RM}^0$ to cancel the interaction term of \eq{eq:rssLp}, we obtain 
\[
Z_L= \frac{1}{Z_{\rm {\it L\,}RM}^0}\lim_{\epsilon\rightarrow +0} \int d\Phi \int_{V_M} dM\, e^{S_{L\Phi M}}, 
\]
wh{}ere
\s[
S_{L\Phi M}^{}:=&i \G_{\kappa \kappa'} \Phi_{l \kappa}^{}\cdot \Phi_{l \kappa'}^{} 
+\sqrt{\epsilon} \tG_{\kappa \kappa'} \Phi_{l \kappa}^{}\cdot \Phi_{l \kappa'}^{}  
-i (p-1) Y \cdot \ss \G_{\kappa \kappa'} \Phi_{l \kappa}^{}
 \Phi_{l \kappa'}^{} v_l^{p-2} \se \\
&\ -i \frac{p-1}{\sqrt{\alpha}} \ss M_l v_l^{p-2} \se \cdot \tIperp \cdot \ss \G_{\kappa \kappa'} \Phi_{l'\kappa}^{}
 \Phi_{l' \kappa'}^{} v_{l'}^{p-2} \se 
- \ss M_l v_l^{p-2} \se \cdot \tIperp \cdot \ss M_{l'} v_{l'}^{p-2} \se,
\label{eq:rsslpm}
\s]
and 
\[
Z_{\rm {\it L\,}RM}^0:= \int_{V_M} dM \exp \left[-\ss M_l v_l^{p-2} \se \cdot \tIperp \cdot \ss  M_{l'} v_{l'}^{p-2} \se \right].
\]

An important fact about the expression \eq{eq:rsslpm} is that it has the same form as \eq{eq:detexp}. 
Here each $\Phi_l$ has its own $J_l(M,\{v\})$, which is given by
\[
J_l(M,\{v\})=I_N-i (p-1) g_{mn} \gtinv{mn}{m'n'} \ss v_{m'} v_{n'}^{p-1} \se \cdot  v_l^{p-2} -i \frac{p-1}{\sqrt{\alpha}} \ss M_{l'} v_{l'}^{p-2} \se \cdot \tIperp \cdot v_{l}^{p-2}.
\]
Then,
\[
Z_L^{}=\frac{1}{Z_{\rm {\it L\,}RM}^0} \int_{V_M} dM\, \left( \prod_{l=1}^L \left| \det J_l(M,\{v\}) \right| \right)
\exp \left[- \ss M_l v_l^{p-2} \se \cdot \tIperp \cdot \ss M_{l'} v_{l'}^{p-2} \se \right].
\label{eq:rsLmat}
\] 

\subsection{Large-$N$ limit}
\label{sec:rslargenL}
In this subsection, we consider the large-$N$ limit of the joint distributions. We assume the scaling
$v_l=N^{-1/(2(p-2))} \tilde v_l$, where $\tilde v_l$ is kept finite in $N\rightarrow \infty$, as in Section~\ref{sec:rslargen}.
We keep $L$ fixed in the limit. 
We take the same procedure of computations using the Schwinger-Dyson equations to compute $Z_{L}^{}$.

The dynamical variables of $Z_{L}^{}$ are $\Phi^{}_{l\kappa}$. These variables contain some components
belonging to $V_\parallel$. However, in the large-$N$ limit, the degrees of freedom of
$\Phi^{}_{l\kappa}$ will be dominated by the components belonging to $V_{\perp}$. 
Therefore, we replace $\Phi_{l\kappa}$ with $\pvp_{l\kappa} \in V_\perp$. 

Let us assume that the supersymmetry \eq{eq:rssusyL} and the discrete symmetry \eq{eq:rsintphiL} are not violated
in the large-$N$ limit. Note also that the orthogonal symmetry (originating from \eq{eq:rsortho}) 
holds in $V_\perp$.  Then, as proven in \ref{app:susytwo}, the two-field correlations are restricted to
have the form,
\[
\langle \pvp_{lmi} \pvp_{l'nj} \rangle =\delta_{ll'} K^{-1}_{mn} Q^{(l)}_{ij} I^\perp
\label{eq:rsassqL}
\]
with 
\[
\ql{l}:=
\left(
\begin{matrix}
\ql{l}_{1} & \ql{l}_2 \\
\ql{l}_2 & \ql{l}_1
\end{matrix}
\right).
\label{eq:theformofQL}
\]

An important property of \eq{eq:rsassqL} is that it is diagonal with respect to the indices $l,l'$. This 
simplifies the large-$N$ computation of the interaction term in \eq{eq:rssLp}: 
\[
&\left \langle  \ss \G_{\kappa \kappa'} \pvp_{l \kappa} \pvp_{l \kappa'} v_l^{p-2} \se \cdot \tIperp \cdot
\ss G_{\kappa'' \kappa'''} \pvp_{l' \kappa''} \pvp_{l' \kappa'''} v_{l'}^{p-2} \se \right\rangle \cr
&\ \ \ =\frac{2}{p(p-1)} 
\ss v_l^{p-2} \se \cdot \ss v_{l'}^{p-2} \se 
\left\langle
\ss \G_{\kappa \kappa'} \pvp_{l \kappa} \pvp_{l \kappa'}\se \cdot \ss G_{\kappa'' \kappa'''} \pvp_{l' \kappa''} \pvp_{l' \kappa'''} \se
\right\rangle \cr
&\ \ \ \simeq \frac{2 \gq{p-2}{l}{l'}}{p(p-1)}(-1)^{\underline{\kappa'}\,\underline{\kappa''}} \G_{\kappa \kappa'}G_{\kappa'' \kappa'''}  \langle \pvp_{l \kappa} \cdot \pvp_{l' \kappa''}\rangle \langle \pvp_{l \kappa'}\cdot \pvp_{l' \kappa'''}\rangle \cr
&\ \ \ = \frac{2 g_{ll}^{(p-2)}}{p(p-1)}(-1)^{\underline{\kappa'}\,\underline{\kappa''}} \G_{\kappa \kappa'}G_{\kappa'' \kappa'''}  \langle \pvp_{l \kappa} \cdot \pvp_{l \kappa''}\rangle \langle \pvp_{l \kappa'}\cdot \pvp_{l \kappa'''}\rangle,
\] 
where, from the second line to the third, we have taken the main large-$N$ contributions of the four-field correlations in terms of 
the two-field correlations (See the footnote in \ref{app:SD}), and, from the third line to the last, 
we have used the fact that $\langle \pvp_{l \kappa} \cdot \pvp_{l' \kappa'}\rangle$ is diagonal with respect to $l,l'$.
Therefore the interaction term is just a sum of each sector of $\pvp_l$.
Since the quadratic terms of \eq{eq:rssLp} are also so, we conclude that the large-$N$ effective action (including
the logarithmic term) is just given by the sum of each sector labeled by $l$. This concludes that 
the large-$N$ asymptotics of $\log Z_L$ is just given by the sum of each sector, which is given 
by \eq{eq:zperplargen}.
It is similarly convenient to introduce $\bar v_l:=(2 (p-1) N/(p\alpha))^{1/(2(p-2))} v_l$. 
Then, by combining with the large-$N$ asymptotics of the overall factors in \eq{eq:rscoeffL} and \eq{eq:rhobarL}, 
we obtain
\s[
\bar\rho_L[\{\bar v\}']=\exp\bigg[N\bigg(& \frac{L}{2} \log [p-1]+ 
\frac{1}{2} \log \left[ 
\frac{\det \bar g }{\det \bgq{p-1}{}{}}
\right]  -\frac{2(p-1)}{p}  \bar g \cdot \bgtinv{}{} \cdot \bar g \\
&\hspace{.5cm}+\sum_{l=1}^L 
{\rm Re} \left[
 \frac{1-\sqrt{1-\bar g_{ll}^{(p-2)}}}{\bar g_{ll}^{(p-2)}} -\log \left[ \frac{1-\sqrt{1-\bar g_{ll}^{(p-2)}}}{\bar g_{ll}^{(p-2)}} \right] 
\right]
\bigg)
+o(N)\bigg],
\label{eq:rsrholargen}
\s]
where $\bar g,\bgq{q}{}{},\bgtinv{}{}$ are defined by performing the replacement $v_l\rightarrow \bar v_l$ 
in $g,\gq{q}{}{},\gtinv{}{}$, respectively. Here $\det \bar g$ in the logarithm has come from 
the overall factor in \eq{eq:rhobarL} by using the equation $\prod_{l=1}^L \bar v_{ll}^2=\det \bar g$, which particularly holds for the parameterization \eq{eq:volgauge}.
\eq{eq:rsrholargen} should hold in any other redundancy-free parameterizations 
than \eq{eq:volgauge}.
 
\section{Joint distributions of complex eigenvectors of complex symmetric random tensor} 
\label{sec:cs}
In this section we extend the analysis of the previous sections to the joint distributions of 
the complex eigenvectors $v_l\in \mathbb{C}^N\ (l=1,2,\cdots,L,\ L\leq N)$ 
of the complex symmetric random tensor $T$. 
The mean distribution can be obtained by setting $L=1$ of the result.
The procedure of computations is similar to the real case with some extensions of the 
conventions/notations to the complex case. 

\subsection{Notations}
\label{sec:cshighernot}
We basically inherit the conventions and the short-hand notations listed in Sections~\ref{sec:rsnotation} and
\ref{sec:rshighernot}
by extending some of the definitions to complex numbers. In the text we may attach $^{\rm c}$ to quantities
to avoid confusions, 
if they contain major structure differences from the corresponding quantities of the real case.

\begin{itemize}
\item
Complex conjugation is denoted by attaching $^*$. For instance $x^*$ denotes the complex conjugate of a vector $x$.

\item
The complex eigenvector equation we consider has a unitary group symmetry \eq{eq:csunitary}. 
We assign $V$ as the index vector space whose invariant transformation is the unitary group. 
We also consider $V^*$ whose invariant transformation is the complex conjugate of the above. 
We distinguish the vectors belonging to either of them by attaching $^*$.
For instance, $x\in V$ and $x^*\in V^*$. The same rule applies to tensors, such as $T\in V^{\otimes p}$ and $T^*\in V^{* \otimes p}$. 

\item
The vector subspace $V_\parallel$ is the complex linear span of the eigenvectors $v_l\ (l=1,2,\cdots,L)$. 
We assume $v_l$ are linearly independent. 
This is a reasonable assumption for $L \leq N$ for an arbitrarily given $T$.
$V_\perp$ is the vector subspace transverse to $V_\parallel$: $v^*\cdot v'=0$ for ${}^\forall v\in V_\parallel, {}^\forall v'\in V_\perp$. 
We denote an arbitrary set of linearly independent vectors spanning $V_\perp$ as $v^\perp_m\ (m=1,2,\cdots,N-L)$. 
$V=V_\parallel \oplus V_\perp$.

\item
The real and imaginary parts of a complex number $x$ are denoted by $\re[x]$ and $\im[x]$, respectively. 
We may also use simpler notations, $x_{\rm R}$ and $x_{\rm I}$, respectively.

\item $T$ denotes a complex symmetric order-$p$ dimension-$N$ tensor, 
whose components $T_{a_1a_2\cdots a_p}\ (a_i=1,2,\cdots,N)$
satisfy $T_{a_1a_2\cdots a_p} \in \mathbb{C}$ and $T_{a_1a_2\cdots a_p}=T_{a_{\sigma[1]}a_{\sigma[2]}\cdots a_{\sigma[p]}}$ 
with $\sigma$ being arbitrary permutations of $\{ 1,2, \cdots, p\}$. The order of the tensor is restricted to $p\geq 3$.  

\item The norm of a vector $v$ is defined by $|v|:=\sqrt{v^*_a v_a}$.

\item The delta function of a complex vector $v$ is defined by $\delta [v]:= \prod_{a=1}^N \delta[v_{a{\rm R}}] \delta[v_{a{\rm I}}]$, i.e., the product of the delta functions of the real and imaginary parts of each component.

\item Dot $\cdot$ between two vectors $x^* \in V^*$ and $y \in V$ represents a contraction $x^* \cdot y:=x^*_a y_a$, and similarly for tensors, 
$X^*\cdot Y:=X^*_{a_1a_2\cdots a_p} Y_{a_1a_2\cdots a_p}$. When the two tensors are the same, we may write
$X^*\cdot X=:|X|^2$.  We also use $\cdot$ for a partial contraction like 
$(X^*\cdot Y)_{a_1a_2\cdots a_p}:=X^*_{a_1a_2\cdots a_p b_1 b_2\cdots b_q} Y_{b_1b_2\cdots b_q}$
or $(X^*\cdot Y)_{a_1a_2\cdots a_p}:=X^*_{b_1 b_2\cdots b_q} Y_{b_1b_2\cdots b_q a_1a_2\cdots a_p  }$.
Note that contractions are always performed between indices belonging to $V$ and $V^*$, 
which preserve the unitary group symmetry.

\item
Dot $\cdot$ may also be used to represent a contraction over $V^*\oplus V$ between doubled variables.
For instance, for $(x^*,x)$ and $(y,y^*)$ $(x,y \in V,\, x^*,y^*\in V^*)$, we have 
$(x^*,x)\cdot (y,y^*):=x^*\cdot y+x\cdot y^*$. This notation applies to tensor contractions as well. 
In the text such dots appear for instance in \eq{eq:csdetjc}.

\item 
Star $\star$ is used to denote the usual matrix product, $(X \star Y)_{ab}:=X_{ac} Y_{cb}$ to avoid confusions with 
the dot $\cdot$ product above. 
$\star$ is assumed to be taken first than the other operations.  In the text it appears for instance in \eq{eq:csdetjc}.

\item The integration measure for a complex $N$-dimensional vector variable $\lambda$ is defined by $d \lambda
:= \prod_{a=1}^N d\lambda_{a{\rm R}} d\lambda_{a{\rm I}}$, and 
the integration region is over $\mathbb{R}^{2N}$,
unless otherwise stated.  Accordingly, we have the following formula,
\[
\delta[f]=\frac{1}{\pi^{2N}} \int d\lambda \, e^{i \,\lambda^* \cdot f+i \, \lambda \cdot f^*}
\label{eq:csdelta}
\]
for a complex vector $f$, and 
\[
\frac{1}{\pi^N}\int d\lambda \, e^{-\lambda^* \cdot A \cdot \lambda}=\frac{1}{\det A}
\]
for an $N$-dimensional positive-definite matrix $A$. 

\item 
Metrics
\s[
&g_{mn}:=v^*_m \cdot v_n, \\
&g^\perp_{mn}:=v_m^{\perp*} \cdot v_n^\perp, \\
&\gq{q}{m}{n}:= v_m^{*q} \cdot v_n^{q}=(g_{mn})^{q}, \\
&\gt{mn}{m'n'}:=\ss v^*_m v_n^{* \, p-1}\se \cdot \ss v_{m'} v_{n'}^{p-1} \se=\frac{1}{p} \left( 
g_{mm'} \gq{p-1}{n}{n'}+(p-1) g_{mn'} g_{nm'} \gq{p-2}{n}{n'} \right). \\
&\gtp{mn}{m'n'}:=\ss v^{*\perp}_m v_n^{*\,p-1}\se \cdot \ss v^\perp_{m'} v_{n'}^{p-1}\se=\frac{1}{p} g^\perp_{mm'} \gq{p-1}{n}{n'}.
\s]
Note that the metrics are hermite.  
\item
Projectors on symmetric order-$p$ tensors
\s[
&\Ipara:=\ss v_mv_n^{p-1} \se \gtinv{mn}{m'n'} \ss v^*_{m'} v_{n'} ^{*\,p-1}\se, \\
&\Iperp:=I-\Ipara, \\
&\tIpara:=\ss v^\perp_mv_n^{p-1} \se \gtpinv{mn}{m'n'} \ss v^{\perp*}_{m'} v_{n'} ^{*\,p-1}\se, \\
&\tIperp:=I-\Ipara-\tIpara.
\s] 
$\gtinv{}{}$ and $\gtpinv{}{}$ denote the inverse matrices of $\gt{}{}$ and $\gtp{}{}$, respectively, 
where they are regarded as matrices with row index $mn$ and column index $m'n'$.  
$I$ denotes an identity operation, and the projectors operate, for example, as $I\cdot T=T$ and 
$\Ipara \cdot T=\ss v_mv_n^{p-1} \se \gtinv{mn}{m'n'} \ss v^*_{m'} v_{n'} ^{*\,p-1}\se\cdot T$
on an arbitrary order-$p$ symmetric tensor $T$. Hence $\Ipara$ and $\tIpara$ are the projections on 
the symmetric tensor subspaces spanned by $\ss v_mv_n^{p-1} \se\ (m,n=1,2,\cdots,L)$ and $\ss v^\perp_m v_n^{p-1} \se\ (m=1,2,\cdots,N-L, n=1,2,\cdots,L)$, respectively. 

\end{itemize}

\subsection{Quantum field theoretical expression}
The complex extension of the real symmetric tensor eigen problem is not unique and 
there are some variations \cite{Kent-Dobias:2020egr,Kent-Dobias2,Majumder:2024ntn}. 
They can be distinguished by the Lie-group symmetries the eigenvector equations have \cite{Majumder:2024ntn}.  
In this paper we consider
\[
T\cdot v^{*\,p-1}=v,
\label{eq:cseg}
\]
which has the symmetry of the unitary group transformation,
\s[
T_{a_1a_2 \cdots a_p}&\rightarrow U_{a_1 a_1'}U_{a_2 a_2'}\cdots U_{a_p a_p'} T_{a'_1a'_2 \cdots a'_p},\\
v_{a}&\rightarrow U_{aa'} v_{a'},
\label{eq:csunitary}
\s]
where $U$ is an arbitrary $N$-dimensional unitary matrix. A reason for selecting this extension among all
is that the asymptotics of the mean distribution has already been shown to have the universal 
form \cite{Delporte:2025tjp}.

We will compute the joint distributions of the eigenvectors $v_l\ (l=1,2,\cdots,L)$.  
The process of computations is very much in parallel with that of the real case,
but with some detailed modifications. Our goal is to compute a quantity which has the same 
appearance as \eq{eq:rsrhoLdef}, but with complex variables: 
\[
\rhocL[\{v\}]:=\left\langle \prod_{l=1}^L \rhoc[v_l,T] \right\rangle_T,
\]
where $\{v\}$ denotes the set of the complex eigenvectors, $\langle \cdot \rangle_T$ denotes the average over 
the complex symmetric random tensor $T$, and  
\[
\rhoc[v,T]&:=| \det \Jc[v,T] |\, \delta [f[v,T]].
\]
Here 
\[
&f[v,T]:=v-T\cdot v^{*\,p-1}, \\
&\Jc[v,T]:=
\left(
\begin{matrix}
\frac{\partial f[v,T]}{\partial v} & \frac{\partial f[v,T]^*}{\partial v} \\
\frac{\partial f[v,T]}{\partial v^*} & \frac{\partial f[v,T]^*}{\partial v^*} 
\end{matrix}
\right), 
\]
where $\Jc[v,T]$ is a complex $2N$-dimensional matrix composed of $N$-dimensional sub-matrices
with components\footnote{$|\det \Jc|$ indeed takes the same value as 
\[
\left | \det 
\left(
\begin{matrix}
\frac{\partial f_{\rm R}}{\partial v_{\rm R}} & \frac{\partial f_{\rm I}}{\partial v_{\rm R}} \\
\frac{\partial f_{\rm R}}{\partial v_{\rm I}} & \frac{\partial f_{\rm I}}{\partial v_{\rm I}} 
\end{matrix}
\right)\right |.
\]
} ,
\[
\left(\frac{\partial f[v,T]}{\partial v}\right)_{ab}:=\frac{\partial f[v,T]_b}{\partial v_a}, \hbox{ and so on}.
\]

As explained in detail in \ref{app:csdetj}, $|\det \Jc|$ can be expressed as
\[
|\det \Jc|=\lim_{\epsilon\rightarrow +0} \frac{1}{\pi^{2N}} \int d\Phic\, 
\exp \left[ 
i \G_{\kappa \kappa'}\, \Phic_\kappa \Phic_{\kappa'} \cdot \bk\star \Jc 
 +\sqrt{\epsilon} \, \tG_{\kappa \kappa'} \Phic_\kappa \Phic_{\kappa'} \cdot  \bk 
 \right],
 \label{eq:csdetjc}
\]
where the superfields are $\Phic_{1i}:=(\bpsi_i^*,\bpsi_i),\Phic_{2i}:=(\psi^*_i,\psi_i),\Phic_{3i}:=(\phi_i^*,\phi_i)$,
$\star$ denotes the matrix product $(\bk\star \Jc)_{ab} =\bk_{ac} \Jc_{cb}$, and $\bk$ is defined in \eq{eq:kbar}.
Note that the components of $\Phic$ are doubled to $V^*\oplus V$, compared to the real case.  
Accordingly, the dot $\cdot$ here represents contractions over $V^*\oplus V$.
Note that  
\[
\bk \star \Jc[v,T]=\left(
\begin{matrix}
-(p-1) \, T\cdot v^{*\,p-2} & I_N \\
I_N & -(p-1)\, T^*\cdot v^{p-2} 
\end{matrix}
\right),
\label{eq:cskj}
\]
which is symmetric, being consistent with the symmetry of $\G_{\kappa \kappa'} \Phic_\kappa \Phic_{\kappa'}$.
By using \eq{eq:csdelta} and \eq{eq:csdetjc}, we obtain the quantum field theoretical expression,
\[
\rhoc_L[\{v\}]
=\lim_{\epsilon\rightarrow +0} \frac{1}{A \pi^{4N L}} \int dT d\lambda d\Phic \, e^{\Sc_L}
\label{eq:csrhoqft}
\]
with
\s[
\Sc_L: =&- \alpha\abs2{T}  +i \lambda_l^* \cdot  f[v_l,T]+i \lambda_l \cdot  f[v_l,T]^* \\
&\ \ \ \ 
+i \G_{\kappa \kappa'}\, \Phic_{l \kappa}  \Phic_{l \kappa'} \cdot \bk \star \Jc[v_l,T] 
 +\sqrt{\epsilon} \, \tG_{\kappa \kappa'} \Phic_{l \kappa}  \Phic_{l\kappa'}\cdot  \bk,
\label{eq:cssmulti}
\s]
and $A:=\int dT\, \exp [-\alpha \abs2{T}]$.
The supersymmetry \eq{eq:rssusyL} and the exchange symmetry \eq{eq:rsintphiL} hold similarly.

\subsection{Integration over $T$ and $\lambda$}
The computations proceed similarly to the real case. Let us first integrate over $T$. 
The terms containing $T$ in \eq{eq:cssmulti} are given by 
\s[
\Sc_{LT}:=&-\alpha\abs2{T} -i \,T\cdot  \ss \lambda_l^*  v_l^{*\, p-1}\se -i \,T^*\cdot  \ss \lambda_l  v_l^{p-1} \se \\
&\ \ \ \ - i (p-1) \G_{\kappa \kappa'} \Phic_{l \kappa}\Phic_{l \kappa'} \cdot 
\left(  
\begin{matrix}
T\cdot v_l^{*\,p-2} & 0 \\
0 & T^*v_l^{p-2} 
\end{matrix}
\right)
\label{eq:csslt}.
\s]
To remove the non-interacting part of $T$ with $\Phic$, as carried out in Section~\ref{sec:rsinttlamL} for the real case, 
we apply the projection $\Ipara$:
\[
T=(\Ipara+\tIpara+\tIperp) \cdot T=Y+  \tIpara  \cdot T+\tIperp  \cdot T, 
\label{eq:cstproj}
\] 
where we have used the eigenvector equation \eq{eq:cseg} to obtain
\[
Y:=\Ipara\cdot T=\ss v_m v_n^{p-1} \se \gtinv{mn}{m'n'} g_{m'n'}.
\]
Putting \eq{eq:cstproj} into \eq{eq:csslt}, we obtain
\s[
\Sc_{LT}=&-i (p-1)  \GPPc \cdot 
\mat2{Y\cdot v_l^{*\, p-2}  & 0 }{0 & Y^*\cdot v_l^{p-2}}
-\alpha \abs2{T} \\
&-i \left(  
\lambda_l^* v_l^{*\,p-1}  +(p-1)  \GPPc v_l^{*\,p-2} \cdot  \mat2{\tIpara+\tIperp &0}{0&0} 
\right)\cdot T \\
&-i \, T^*\cdot \left( \lambda_l v_l^{p-1}+(p-1)  \mat2{0&0}{0& \tIpara+\tIperp}  \cdot \GPPc v_l^{p-2}\right).
\label{eq:csscltmid}
\s]

As in \eq{eq:sllam} for the real case, it is now convenient to carry out a splitting 
$\lambda_l=\lambda_l^\parallel+\lambda_l^\perp\ (\lambda_l^\parallel\in V_\parallel,\lambda_l^\perp\in V_\perp)$.
We can remove the projection $\tIpara$ from \eq{eq:csscltmid} by shifting $\lambda_l^\perp$ to absorb it.
Then, by integrating over $T$, we obtain
\s[
\frac{1}{A} \int &dT e^{\Sc_{LT}}=\exp \Bigg[-i (p-1)  \GPPc \cdot 
\mat2{Y\cdot v_l^{*\, p-2}  & 0 }{0 & Y^*\cdot v_l^{p-2}} \\
&-\frac{1}{\alpha} \ss \lambda^{\parallel *}_l v_l^{*\,p-1} \se \cdot \ss \lambda_{l'}^\parallel v_{l'}^{p-1} \se
-\frac{1}{\alpha} \ss \lambda^{\perp *}_l v_l^{*\,p-1} \se \cdot \ss \lambda_{l'}^\perp v_{l'}^{p-1} \se \\
&-\frac{(p-1)^2}{\alpha}\left(  \GPPc v_l^{*\, p-2} \cdot \mat2{I_N&0}{0&0} \right) \cdot \tIperp \cdot \left( \mat2{0&0}{0&I_N} \cdot
\GPPcd v_{l'}^{ p-2} \right)
\Bigg].
\label{eq:csresofTint}
\s]

To perform the integration over $\lambda$, let us collect the terms containing $\lambda$. They are those in \eq{eq:csresofTint} and 
$i \lambda_l^* \cdot v_l+i \lambda \cdot v_l^*$ from \eq{eq:cssmulti}.
The computation can straightforwardly be done, for instance, by expanding $\lambda_l^\parallel=\lambda_{lm} v_m$, and 
we obtain
\[
\rhocL[\{v\}]=\left( \frac{\alpha}{\pi}\right)^{NL} p^{L(N-L)} (\det g)^L (\det \gt{}{})^{-1} (\det \gq{p-1}{}{})^{-(N-L)} 
\exp [-\alpha \, g^* \cdot \gtinv{}{} \cdot g] \ZcL,
\label{eq:csfinalrho}
\]
where $g^* \cdot \gtinv{}{} \cdot g=g_{mn}^* \gtinv{mn}{m'n'} g_{m'n'}$, and 
\[
\ZcL:=\lim_{\epsilon\rightarrow +0} \frac{1}{\pi^{2 NL}} \int d\Phic e^{\Sc_{L\Phi}}
\]
with
\s[
\Sc_{L\phi}:=&i \G_{\kappa\kappa'} \Phic_{l\kappa}\Phic_{l\kappa'}\cdot \bk  +\sqrt{\epsilon}  \tG_{\kappa\kappa'} 
\Phic_{l\kappa}\Phic_{l\kappa'}\cdot \bk
-i (p-1)  \GPPc \cdot \mat2{Y\cdot v_l^{*\, p-2}  & 0 }{0 & Y^*\cdot v_l^{p-2}}\\
&-\frac{(p-1)^2}{\alpha}\left(  \GPPc v_l^{*\, p-2} \cdot \mat2{I_N&0}{0&0}\right) \cdot \tIperp \cdot  
\left(\mat2{0&0}{0&I_N} \cdot \GPPcd v_{l'}^{p-2}\right).
\label{eq:csfinalsc} 
\s]
We see that, comparing with \eq{eq:rscoeffL} of the real case, 
the result \eq{eq:csfinalrho} of the complex case has a form more or less expected from doubling the variables.  

The result \eq{eq:csfinalrho} is invariant under the unitary transformation in $V$. 
A parametrization with no redundancies is derived in \ref{app:volume}:
\[
v_{la}=0 \hbox{ for } a>l,\ v_{ll}>0,
\label{eq:cspara}
\]
where $v_{la}\ (a<l)$ can take any complex numbers and 
$v_{ll}$ any positive numbers, 
and the associated infinitesimal volume is given by
\[
dv':=\prod_{l=1}^L S_{2(N-i+1)-1} v_{ll}^{2(N-i+1)-1} dv_{ll} \prod_{a=1}^{l-1} dv_{la{\rm R}} dv_{la{\rm I}}.
\label{eq:csvol}
\]
Using \eq{eq:cspara} and the above parametrization $\{v\}'$, we obtain
\[
\bar \rho^{\rm c}_L[\{ v\}']=\left( \prod_{l=1}^L S_{2(N-i+1)-1} v_{ll}^{2(N-i+1)-1} \right) \rhocL[\{ v\}].
\label{eq:csrhobar}
\]

\subsection{Random matrix representation of $\ZcL$}
\label{sec:csrandom}
In this subsection we obtain a random matrix representation of $\ZcL$. The derivation is similar to the real case,
so we describe its main points only. 
  
The interaction term in \eq{eq:csfinalsc} can be expressed as
\[
\ss H_l^{{\rm c}*} v_l^{*\,p-2} \se \cdot \tIperp \cdot \ss \Hc_{l'} v_{l'}^{p-2} \se
\]
with\footnote{Note $H^{{\rm c}*}_l=\frac{p-1}{\sqrt{\alpha}} \GPPc\cdot \mat2{I_N&0}{0&0}$.}
\[
\Hc_l:=\frac{p-1}{\sqrt{\alpha}} \GPPc\cdot \mat2{0&0}{0&I_N}.
\]
Therefore, by introducing complex symmetric matrices $M_l$, we consider
\[
Z_{\rm RM}^{\rm c}:=\int_{V^{\rm c}_M} dM \, \exp \left[ 
-\ss M^{*}_l  v_l^{*\,p-2} +i H_l^{{\rm c}*} v_l^{*\,p-2}   \se \cdot \tIperp \cdot \ss M_{l'} v_{l'}^{p-2}   + i \Hc_{l'} v_{l'}^{p-2} \se
\right].
\]
Here the integration region $V_M^{\rm c}$ is a quotient vector space,
\[
V_M^{\rm c} :\cong V_{CS}/K^{\rm c}_{\tIperp\cdot  v^{p-2} },
\label{eq:defofvmc}
\]
where $V_{CS}$ is the vector space of $L$ complex symmetric matrices of dimension $N$, and $K^{\rm c}_{\tIperp\cdot  v^{p-2} }$ 
is a kernel, 
\[
K^{\rm c}_{\tIperp\cdot  v^{p-2} }:=\{(M_1,M_2,\cdots,M_L ) \,
|\,  \tIperp\cdot \ss  M_{l} v_{l}^{p-2} \se=0, (M_1,M_2,\cdots,M_L )\in V_{CS} \}.
\]
Canceling the interaction term in \eq{eq:csfinalsc} by multiplying $1=Z_{\rm RM}^{\rm c}/Z_{\rm RM}^{\rm c\,0}$, 
we obtain
\[
\ZcL=\frac{1}{Z^{{\rm c}\,0}_{\rm RM}} \lim_{\epsilon \rightarrow +0} \int d\Phic \int_{V^{\rm c}_M} dM \, e^{S^{\rm c}_{L\phi M}},
\]
where 
\s[
S^{\rm c}_{L\phi M}:=&i \G_{\kappa\kappa'} \Phic_{l\kappa}\Phic_{l\kappa'}\cdot \bk  +\sqrt{\epsilon}  \tG_{\kappa\kappa'} 
\Phic_{l\kappa}\Phic_{l\kappa'}\cdot \bk
-i (p-1)  \GPPc \cdot \mat2{Y\cdot v_l^{*\, p-2}  & 0 }{0 & Y^*\cdot v_l^{p-2}}\\
&-\frac{i(p-1)}{\sqrt{\alpha}} M^{*}_lv_l^{*\,p-2} \cdot \tIperp \cdot \left(  \mat2{0&0}{0&I_N} \cdot \G_{\kappa \kappa'} \Phic_{l'\kappa}
\Phic_{l'\kappa'} v_{l'}^{p-2} \right)  \\
&-\frac{i(p-1)}{\sqrt{\alpha}}\left( \G_{\kappa \kappa'} \Phic_{l\kappa}\Phic_{l\kappa'} v_{l}^{*\,p-2}\cdot 
\mat2{I_N&0}{0&0} \right) \cdot \tIperp \cdot M_{l'} v_{l'}^{p-2} \\
&-\ss M^{*}_l  v_l^{*\,p-2}   \se \cdot \tIperp \cdot \ss M_{l'} v_{l'}^{p-2} \se, 
\label{eq:csscl}
\s]
and 
\[
Z_{\rm RM}^{{\rm c}\,0}:=\int_{V^{\rm c}_M} dM \, \exp \left[ 
-\ss M^{*}_l  v_l^{*\,p-2}   \se \cdot \tIperp \cdot \ss M_{l'} v_{l'}^{p-2}   \se
\right].
\label{eq:csrandomZ}
\]

\eq{eq:csscl} has the form of the righthand side of \eq{eq:csdetjc} with a $J^c$ for each $\Phic_l$. 
We obtain
\[
\ZcL=\frac{1}{Z_{\rm RM}^{{\rm c}\,0}} \int_{V_M^{\rm c}} dM\, 
\left( \prod_{l=1}^L \left| \det J_l^{\rm c}(M,\{v\}) \right| \right) 
\exp \left[ -\ss M^{*}_l  v_l^{*\,p-2}   \se \cdot \tIperp \cdot \ss M_{l'} v_{l'}^{p-2}   \se \right],
\label{eq:csLmat}
\] 
where
\s[
&J_l^{\rm c}(M,\{v\}) \\
&=\mat2{
I_{N} & -(p-1) \left( Y^*+\frac{1}{\sqrt{\alpha}} M^{*}_{l'} v_{l'}^{*\,p-2} \cdot \tIperp  \right) \cdot v_l^{p-2}}
{-(p-1) v_l^{*\,p-2} \cdot \left( Y+\frac{1}{\sqrt{\alpha}}  \tIperp \cdot M_{l'} v_{l'}^{p-2}  \right)  &I_N}.
\label{eq:csjlc}
\s]

\subsection{$L=1$ case}
In this subsection we take $L=1$ in the result of Section~\ref{sec:csrandom}, 
namely, we more explicitly study the random matrix representation of the mean distribution 
of the complex eigenvectors. 

Let us take an orthonormal basis of $V_\perp$ as $\vp_m\ (m=1,2,\cdots,N-1)$, which satisfy $\vp_m{}^* \cdot \vp_n=\delta_{mn}$.  Then we obtain
\[
\tIperp{}^{L=1}=I-\frac{v^p v^{*\,p}}{|v|^{2p}}-\frac{p}{|v|^{2(p-1)}}\ss \vp_m v^{p-1} \se \ss \vp_m{}^* v^{* \, p-1}\se. 
\label{eq:csprojLeq1}
\]
For $L=1$ there is only one $M$, which is a complex symmetric matrix. It can straightforwardly be seen 
that the projection of \eq{eq:csprojLeq1} to $\ss M v^{p-2}\se$
projects out the components of $M$ which are linear combinations of $v^2$ and ${}^\forall \ss \vp_m v\se$ . Therefore a choice of the integration
region is given by  $V_M^{{\rm c}\,L=1}=\{ M^\perp | M^\perp\cdot v=0\}$. Putting this $M^\perp$ into \eq{eq:csjlc}, we obtain
\[
J^{{\rm c}\,L=1} (M^\perp,v)=\mat2{
I_N & -(p-1) \left( \frac{v^{*\,2}}{|v|^2} +\frac{2 |v|^{2(p-2)}}{\sqrt{\alpha} p (p-1)} M^{\perp*}\right) }
{-(p-1) \left( \frac{v^{2}}{|v|^2} +\frac{2 |v|^{2(p-2)}}{\sqrt{\alpha} p (p-1) }M^{\perp}\right) & I_N}
\]
Then putting this to \eq{eq:csrandomZ} 
and performing a rescaling, $M^\perp\rightarrow \sqrt{2^{-1} p(p-1)}|v|^{-p+2} M$, 
we obtain
\[
Z^{\rm c}_{L=1}=\frac{p(p-2)}{Z^{{\rm c}\,0}_{L=1}} \int_{V^{\rm c}_{N-1}}  dM \det \left(I_{N-1}- \xi M^* \star M \right)
\exp[- M^*\cdot M],
\label{eq:csZcLeq1}
\] 
where $Z^{{\rm c}\,0}_{L=1}:=\int_{V^{\rm c}_{N-1}}  dM \exp[- M^*\cdot M]$, $\xi:=2 (p-1) |v|^{2(p-2)}/(\alpha p)$, 
and $\star$ denotes a matrix product $(M^* \star M)_{ab}=M^*_{ac}M_{cb}$.
Here the integration region $V^{\rm c}_{N-1}$ is over  the $(N-1)$-dimensional complex symmetric matrix.

The complex symmetric matrix $M$ can be diagonalized by the Autonne-Takagi factorizaiton,  $M=UDU^T$, where 
$U$ is a unitary matrix and $D$ is a diagonal matrix with non-negative real entries. 
Accordingly, the matrix model in \eq{eq:csZcLeq1}
can analytically be solved. In this case, after integration over $U$,  
the integration measure over the diagonal entries $\zeta_i\ (i=1,\cdots,N-1)$
is given by $(\prod_{i} \zeta_i d\zeta_i )( \prod_{j<k} |\zeta_j-\zeta_k| |\zeta_j+\zeta_k|)$. 
By changing the variable to $z_i:=\zeta_i^2$,  we obtain
\[
Z^{\rm c}_{L=1}=\frac{p(p-2) \xi^{N-1} e^{z_{N}}  \int_0^\infty dz_1 \cdots dz_{N-1}\, P^{(N)}(z_1,\cdots,z_{N})}
{\int_0^\infty dz_1 \cdots dz_{N-1}\, P^{(N-1)}(z_1,\cdots,z_{N-1})}, 
\label{eq:cszperpexact}
\]
where $z_{N}=1/\xi$ and 
\[
P^{(n)}(z_1,\cdots,z_n):=e^{-\sum_{i=1}^n z_i} \prod_{i<j} |z_i-z_j|.
\]
For the self-containedness of this paper, the explicit functional forms are given in \ref{app:csexplicitI}.

\subsection{Large-$N$ limit}
In this section we discuss the large-$N$ limit of the complex case. We employ a similar procedure as the real case 
in Section~\ref{sec:rslargenL}.
We assume the same scaling $v_l=N^{-1/(2(p-2))} \tilde v_l$, and that the degrees of freedom associated to $V_\parallel$ can be ignored in the large-$N$ limit. Namely, we consider only $\pvpc\in V_\perp$ as $\Phic$.

Let us first note that the supersymmetry and the exchange symmetry hold also in the complex case 
(See the comment below \eq{eq:cssmulti}). 
Also taking into account the unitary symmetry \eq{eq:csunitary}, 
the two field correlation is constrained to have the form, 
\[
\langle \pvpc_{lmi} \pvpc_{l'nj} \rangle =\delta_{ll'} K^{-1}_{mn} Q^{(l)}_{ij} \bk^\perp,
\label{eq:csassqL}
\]
where $\bk^\perp$ denotes $\bk$ restricted to the subspace $V^*_\perp\oplus V_\perp$, 
and $Q^{(l)}$ have the form given in \eq{eq:theformofQL}.
Putting \eq{eq:csassqL} into \eq{eq:csscl}, we obtain 
\[
S_{\rm eff}^{\rm c}=N\left( -4i Q_2-4 \sqrt{\epsilon} Q_1 + 4 \tilde \xi (Q_1^2+Q_2^2) - \log(Q_1^2-Q_2^2) \right),
\]
where $\tilde \xi:=N\xi = (p-1) |\tilde v|^{2(p-2)}/(\alpha p)$. Comparison with \eq{eq:rsseff} shows that the effective action 
is twice that of the real case with a factor of 2 difference of the coupling constant (See \eq{eq:rsdefg}).   
Therefore further analysis of the effective action
proceeds in the same manner. It is also straightforward to take the large-$N$ limit of the overall 
factors in \eq{eq:csfinalrho} and \eq{eq:csrhobar}.
Then, by introducing 
$\bar v_l:=(4(p-1)/(\alpha p))^{1/(2(p-2))} \tilde v_l=(4(p-1)N/(\alpha p))^{1/(2(p-2))} v_l$, we obtain 
the large-$N$ asymptotics, 
\s[
\bar \rho^{\rm c}_L[\{v\}']
=\exp\bigg[2N\bigg(& \frac{L}{2} \log [p-1]+ 
\frac{1}{2} \log \left[ 
\frac{\det \bar g }{\det \bgq{p-1}{}{}}
\right]  -\frac{2(p-1)}{p}  \bar g^* \cdot \bgtinv{}{} \cdot \bar g \\
&\hspace{.5cm}+\sum_{l=1}^L 
{\rm Re} \left[
 \frac{1-\sqrt{1-\bar g_{ll}^{(p-2)}}}{\bar g_{ll}^{(p-2)}} -\log \left[ \frac{1-\sqrt{1-\bar g_{ll}^{(p-2)}}}{\bar g_{ll}^{(p-2)}} \right] 
\right]
\bigg)
+o(N)\bigg],
\label{eq:csrholargen}
\s]
which has the exponent exactly twice that of the real case in \eq{eq:rsrholargen} with straightforward 
replacements of the metrics to the complex ones in Section~\ref{sec:cshighernot}.  

\section{Numerical crosschecks}
\label{sec:numerical}
In this section we perform some numerical crosschecks of our results by Monte Carlo (MC) simulations. A major end is to show the validity of 
the large-$N$ asymptotics of the eigenvector joint distributions, which have been derived under a few plausible
assumptions about the symmetries and the relevant degrees of freedom in large-$N$. 
It is not possible to directly compare the asymptotics with the MC computations of 
the joint distributions of the eigenvectors
of the random tensors, because such MC computations quickly become too time-consuming, as $N$
becomes larger.
Therefore we perform the crosschecks in two steps: 
We first show numerically the validity of the random matrix representations for small $N$ by comparing with 
the joint distributions computed by explicitly solving the eigenvectors of randomly generated
tensors, and then compare the former for very large $N$ with the asymptotics,
where $N$ is taken about several hundreds.  We find convincing agreements.  

Another important conclusion obtained from the crosschecks is that, since the numerical computations contain
no regularizations (equivalent to taking $\epsilon\rightarrow +0$ first), the numerical support 
of the large-$N$ analytic asymptotic formula in Section~\ref{sec:largenmc} implies the commutativity of the 
two limits, $\epsilon\rightarrow +0$ and $N\rightarrow \infty$, independently from the proof 
of \ref{app:commute}.

\subsection{Monte Carlo simulations}
We perform two kinds of MC simulations\footnote{The numerical computations are
performed on a laptop computer.}: (i) 
the direct computations of the joint distributions by
explicitly solving the eigenvectors of randomly generated tensors, and 
(ii) the computations of the random matrix representations by generating randomly the matrices. 

As for (i), a tensor $T$ is generated randomly 
with the Gaussian distribution, and then all the eigenvectors, satisfying \eq{eq:egvec} or \eq{eq:cseg},
 are computed by using the polynomial equation 
solvers of Mathematica\footnote{We used Mathematica 13.}. 
Then the joint distributions of the eigenvectors are computed. 
The procedures of the MC and the data analysis are essentially the same as in the previous studies 
(Some details can be found for instance in \cite{Sasakura:2022iqd}). 
Because of the complication of solving the non-linear polynomial equations for eigenvectors, the computations are rather slow, 
and they are possible only for $N\lesssim 10$ for $p=3$ on a laptop computer.
We also find the speed becomes substantially slow for $p\geq 4$. 
Therefore we restrict ourselves to $p=3, N=4$, considering the overall statistical performance. 

As for (ii), the random matrix representations  \eq{eq:rsLmat} or \eq{eq:csLmat} 
are rewritten more suited for numerical studies 
as \eq{eq:rszltilde} or \eq{eq:csrszltilde} in \ref{app:actualnumerics}, respectively.
They are Gaussian matrix models with certain determinant factors. This can easily be implemented as 
the MC computations of the mean values of the determinant factors with Gaussian random variables.
The computations are much faster than the above MC with tensors, 
and can be performed rather easily with $N\sim$ several hundreds. 
Therefore the results can be well compared with the asymptotics.     

We use the parameterizations \eq{eq:volgauge} and \eq{eq:cspara} of $v$ 
for the real and the complex cases, respectively. 
The sample number in MC is denoted by $\#$samp.

\subsection{Mean distributions}
We first check the mean distributions of the eigenvectors, which correspond to setting $L=1$. 
This case gives good tests for our MC method, 
which uses the polynomial equation solvers of Mathematica to compute all the eigenvectors. 
 
\begin{figure}
\hfil
\includegraphics[width=7cm]{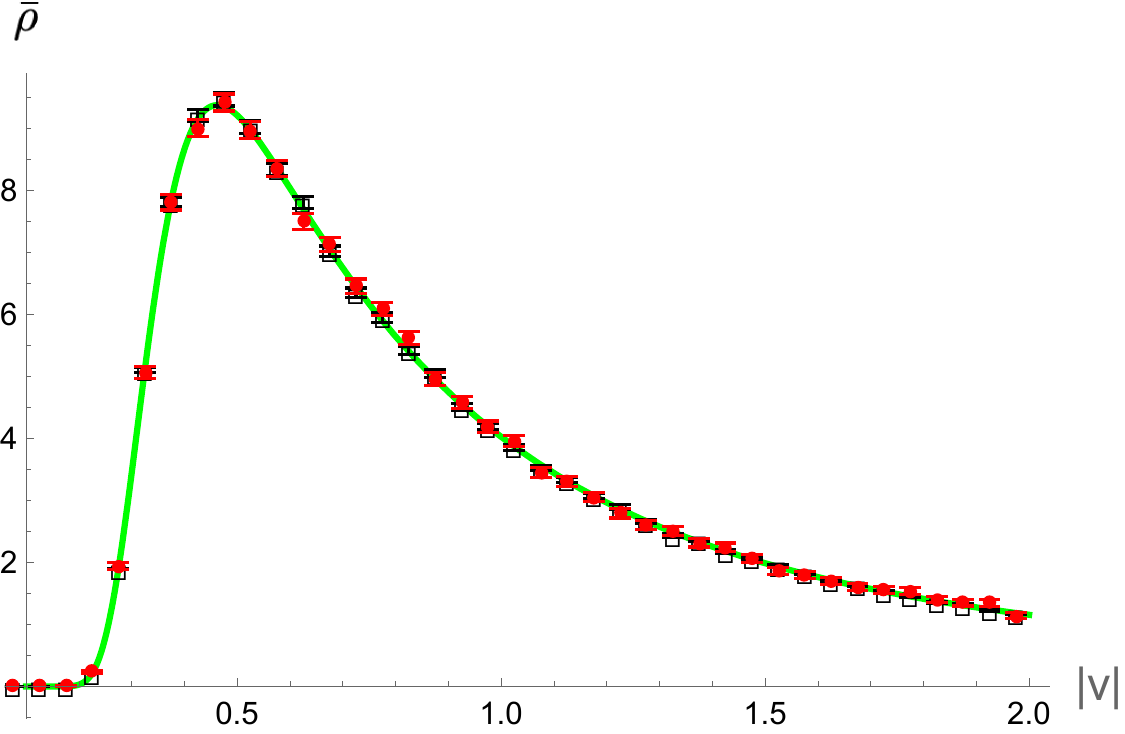}
\hfil
\includegraphics[width=7cm]{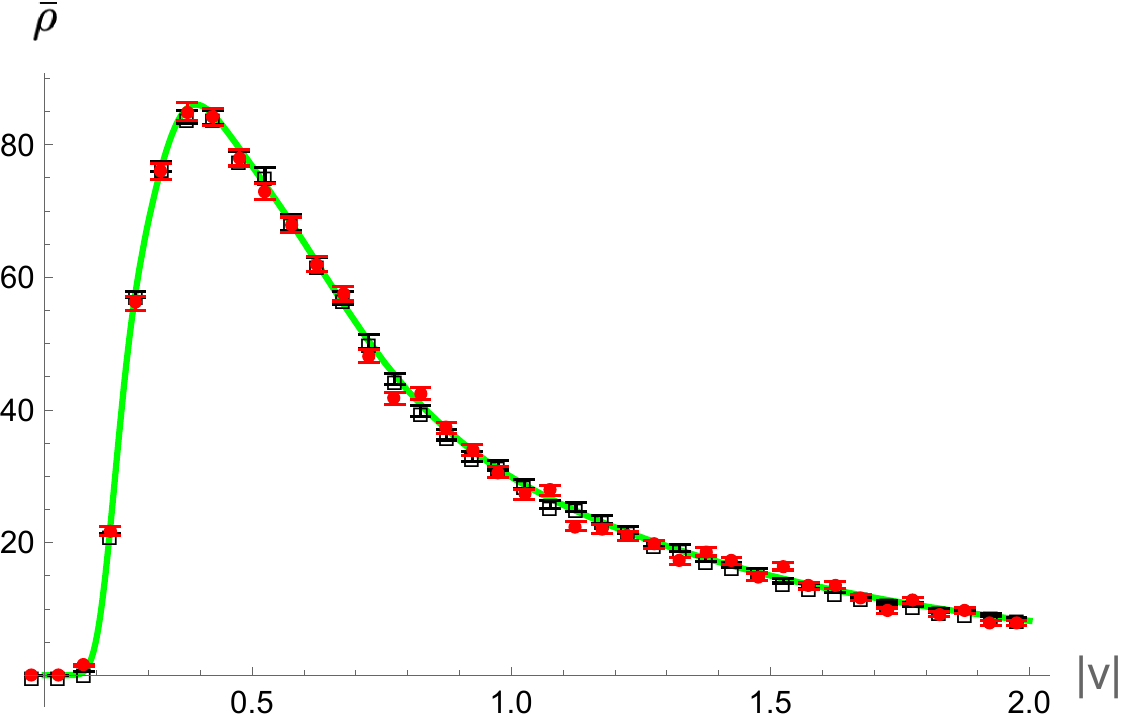}
\hfil
\caption{The mean distributions ($L=1$) for $p=3,N=4,\alpha=1/2$, 
computed by the three different ways (i),(ii), and the analytic expression:
Red points, Black squares, Green solid line, respectively.
Left: Real case. $\#$samp=$10^4$ for both (i) and (ii).  
Right: Complex case.  $\#$samp=$10^3$ for (i) and $10^4$ for (ii).
Many of (ii) are hidden behind (i), because of good agreements.}
\label{fig:one}
\end{figure}

In Figure~\ref{fig:one}, we plot three quantities, (i), (ii), and the exact analytical expressions \eq{eq:rsrhobar} with \eq{eq:rszperpexact} or \eq{eq:csfinalrho} with \eq{eq:csrhobar} and \eq{eq:cszperpexact}, for the real (left) and the complex (right) case, respectively.
They are respectively (i) Red points with statistical error bars, 
(ii) Black squares with statistical error bars, and a green solid line. 
The agreements are very satisfactory. 

\subsection{Joint distributions}
\begin{figure}
\hfil
\includegraphics[width=7cm]{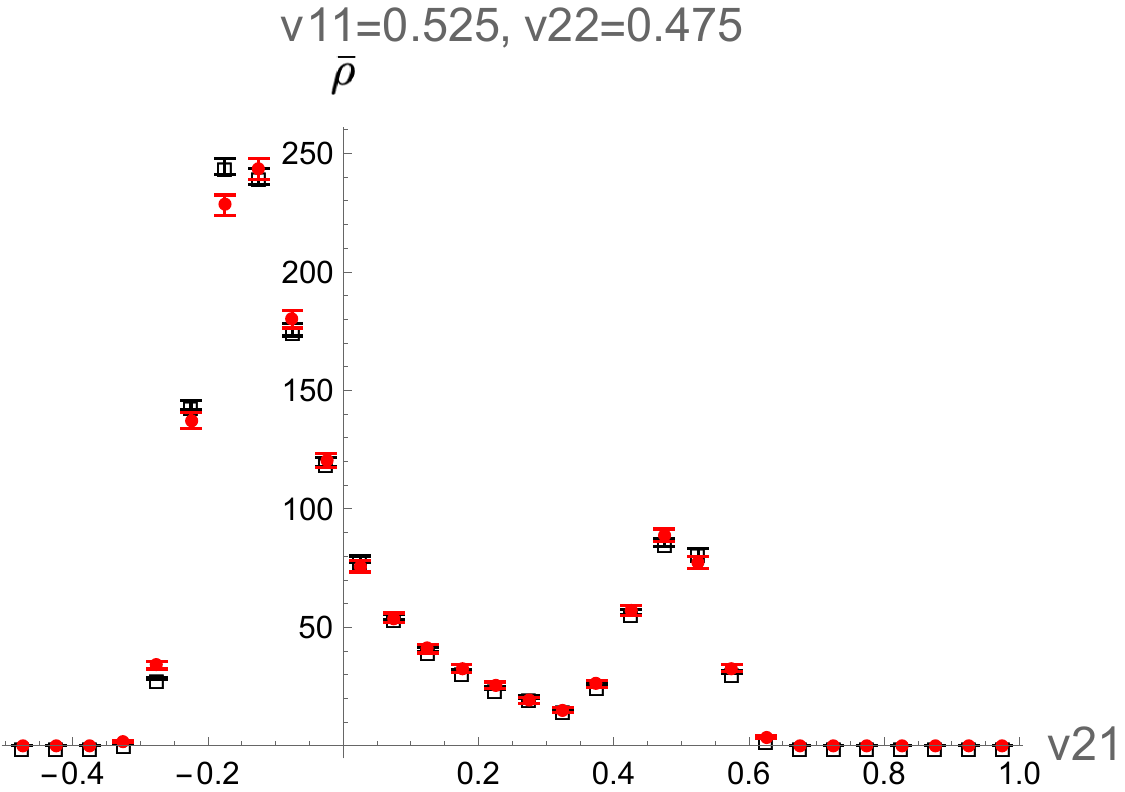}
\hfil
\includegraphics[width=7cm]{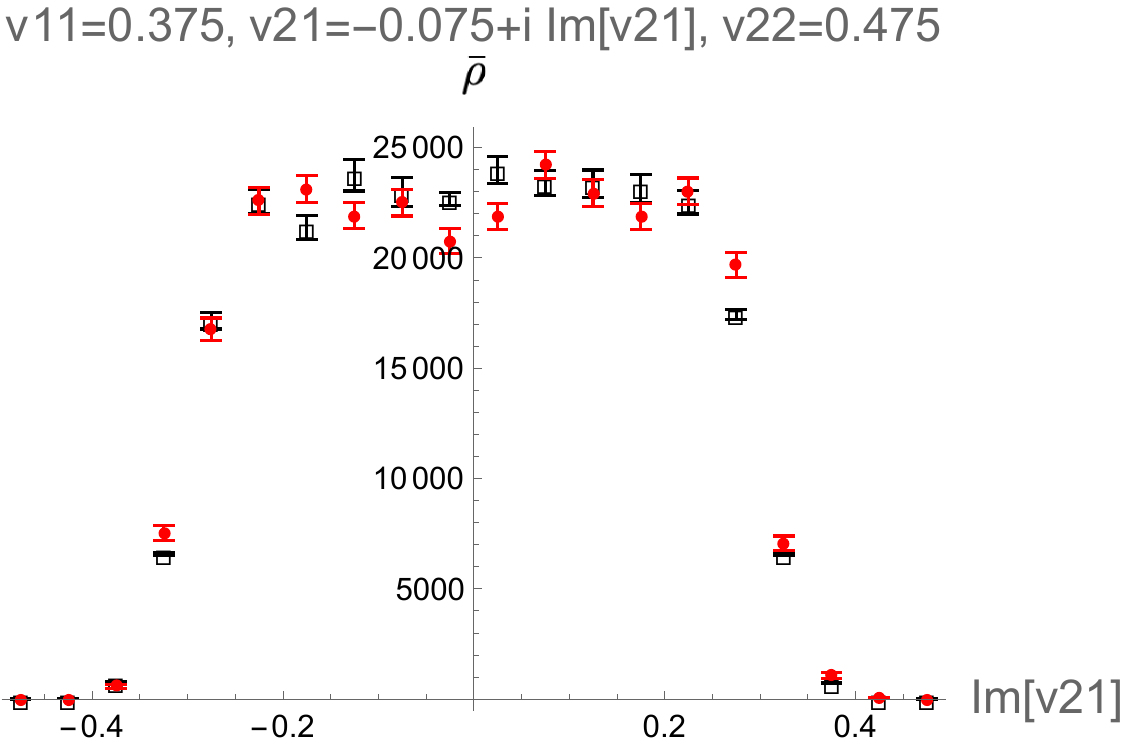}
\hfil
\caption{The joint distributions of two eigenvectors ($L=2$) 
for $p=3,N=4,\alpha=1/2$, computed by the two different methods (i),(ii). 
Left: A real case. $\#$samp=$10^5$ for (i) and $10^4$ for (ii). 
Right: A complex case. $\#$samp=$10^4$ for both (i) and (ii).}
\label{fig:two}
\end{figure}

In this subsection we check the agreement between (i) and (ii) for $L=2,3$, $p=3$, $N=4$.
Because the parameterizations \eq{eq:volgauge} and \eq{eq:cspara} are high dimensional, 
we show them only for some arbitrarily chosen parameter slices. 
Figure~\ref{fig:two} shows such comparisons for $L=2$. 
The real case is shown for certain arbitrarily taken values of $v_{11},v_{22}$ with variable $v_{21}$,
and the complex case with variable $\hbox{Im}[v_{21}]$. 
They show good agreements.

\begin{figure}
\hfil
\includegraphics[width=7cm]{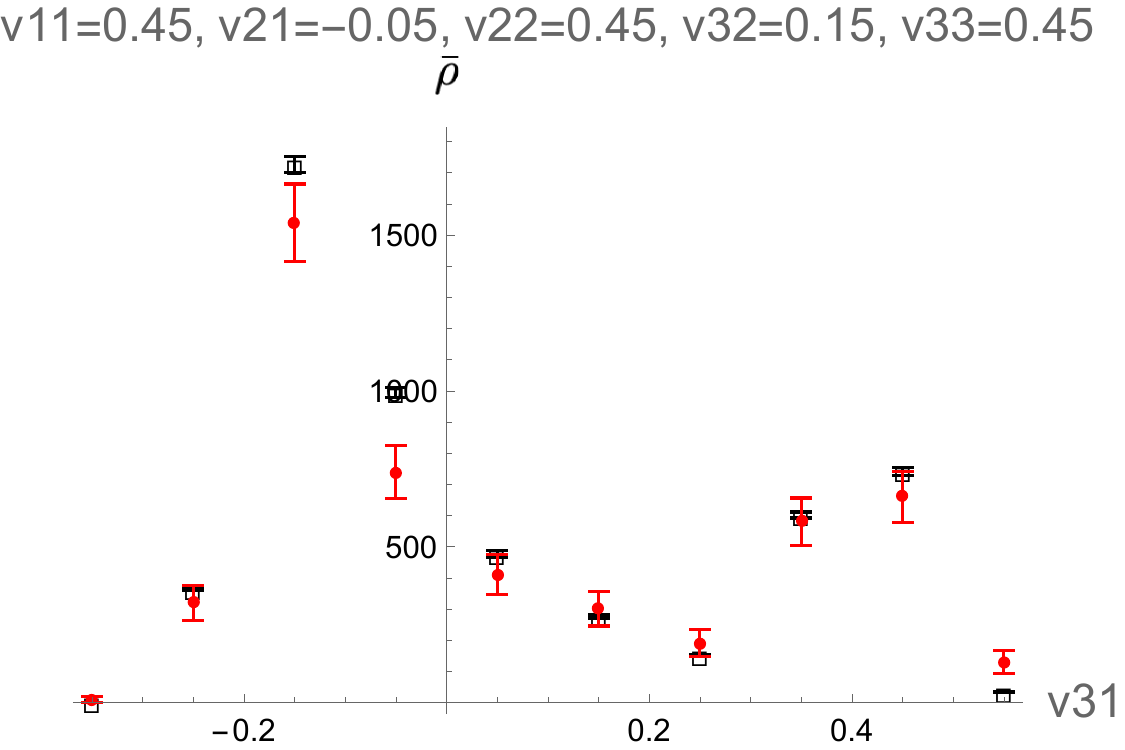}
\hfil
\caption{The joint distributions of three eigenvectors ($L=3$)
for $p=3,N=4,\alpha=1/2$, computed by the two different methods (i),(ii). 
A real case. $\#$samp=$10^5$ for (i) and $10^4$ for (ii). }
\label{fig:three}
\end{figure}

The comparison for $L=3$ for the real case is shown in Figure~\ref{fig:three}, 
and they agree well. The complex case is not shown, 
because the dimension of the parameters is too large for our machine power
to get statistically sensible data. 

\subsection{Large-$N$}
\label{sec:largenmc}
\begin{figure}
\hfil
\includegraphics[width=7cm]{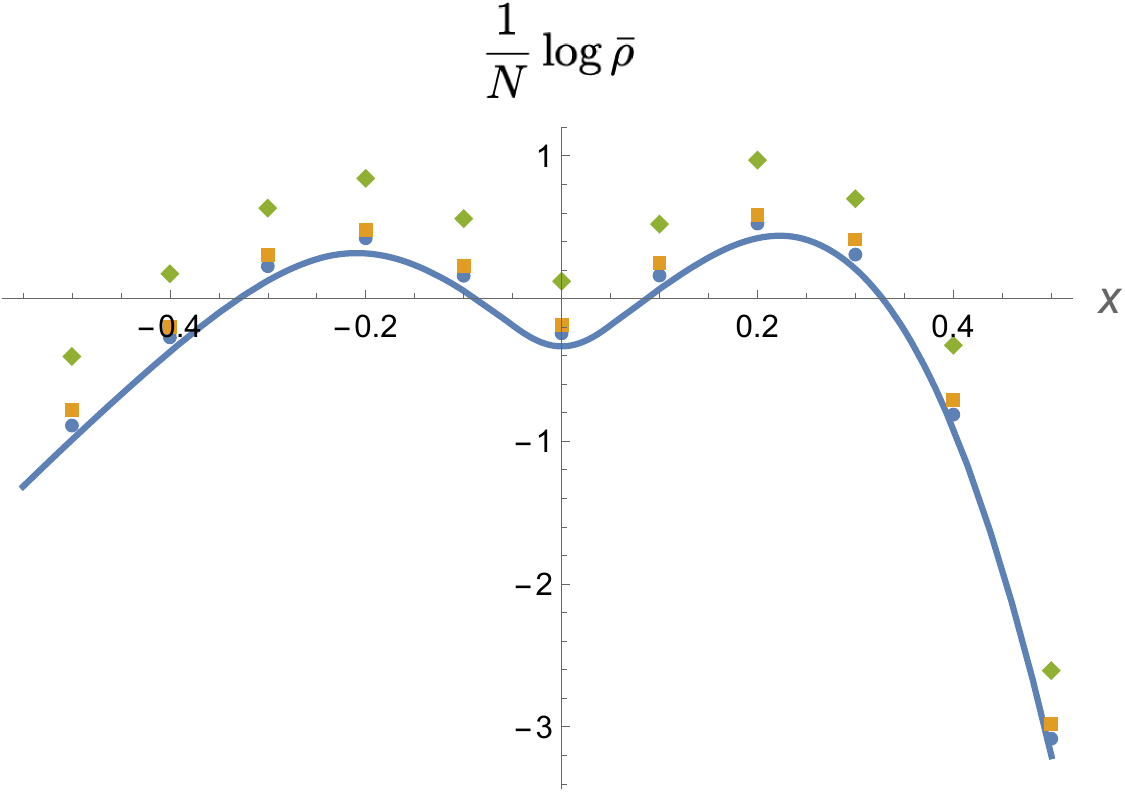}
\hfil
\includegraphics[width=7cm]{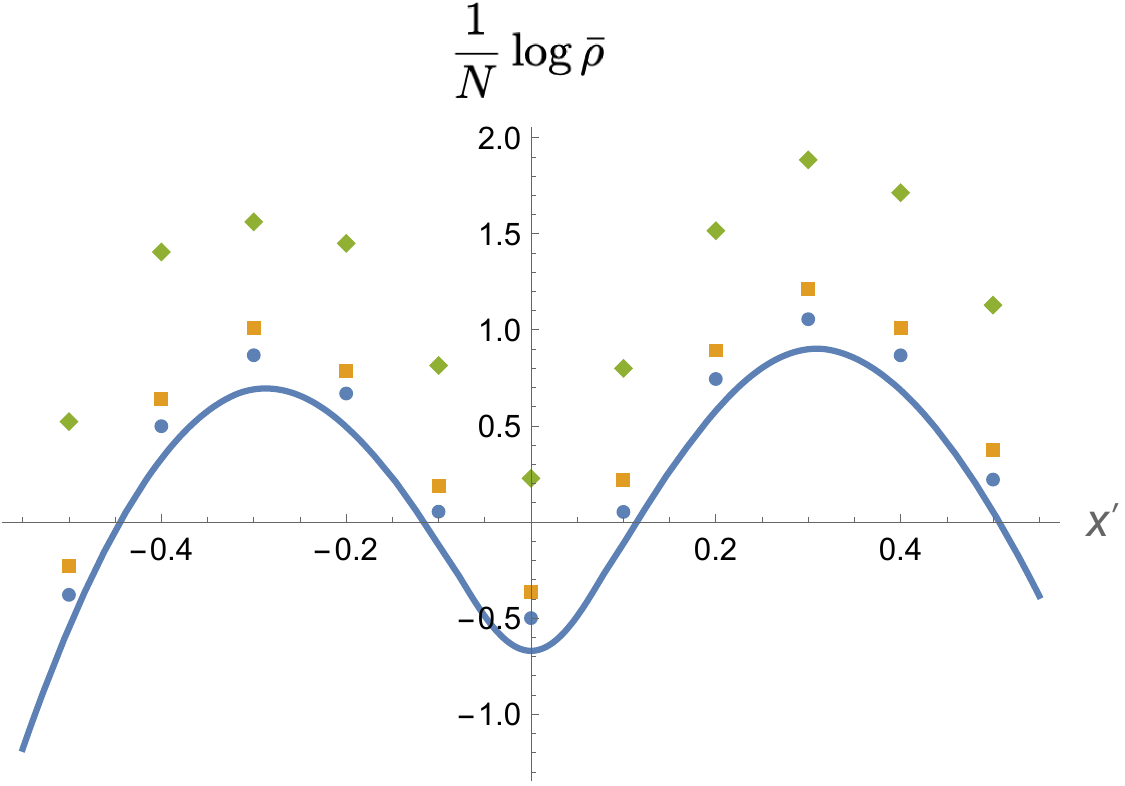}
\hfil
\caption{Comparison of the large-$N$ asymptotics (solid lines) with the random matrix representations 
for $p=5$, $L=4$ and $N=$50 (diamonds), 200 (squares), 400 (dots). 
$\bar v$'s are parameterized as 
$\{ \bar v_{11},\bar v_{21} ,\bar v_{22},\bar v_{31},\bar v_{32} ,\bar v_{33},\bar v_{41},\bar v_{42} ,
\bar v_{43},\bar v_{44}\}=\{y, x, y, -x, -2 x, y, 3 x, 2 x, x, y\}$ and the others are set zero. 
Left: A real case. $y=0.97$ and variable $x$.  Right: A complex case. $y=0.97,\, x=(0.3+0.7 i) x'$ with 
real variable $x'$. }
\label{fig:largeN}
\end{figure}

In this subsection we finally check 
the large-$N$ asymptotic expressions \eq{eq:rsrholargen} and \eq{eq:csrholargen} by
comparing with the random matrix representations. 
In Figure~\ref{fig:largeN} we show an example for each real and complex case. 
$\#$samp=3 for the random matrix
representations\footnote{In fact one sample is enough to get the value itself, but 
a few would be needed to evaluate the statistical error.}.
This $\#$samp is very small, but the errors are even smaller than or similar to the size of the data markers
in the figure,
because the systems become non-fluctuating in the large-$N$ limit (like a thermodynamic limit
with an infinite number of degrees of freedom).
The MC computations can be carried out fast enough even for $N$ of several hundreds. 
As in the figure, we find that the data get closer to the large-$N$ asymptotics, 
as $N$ becomes larger, supporting their validities.      

\section{Summary and future prospects}
In this paper we have studied the joint distributions of arbitrary numbers of eigenvectors of 
the real and complex symmetric random tensors. We have obtained the random matrix representations 
and the large-$N$ asymptotics of the distributions. We have numerically checked the results by 
Monte Carlo computations. 

While the random matrix representations give explicit analytic expressions only for the mean distributions, 
they provide numerical measures to compute the joint distributions much more efficiently than by numerically
solving the tensor eigenvectors of randomly generated tensors. 
Therefore these representations can provide efficient numerical assistance for
further studies of the eigen problems, especially when tensors have large dimensions. 
They also allow the possibilities to apply some matrix model techniques, such as large deviations  \cite{largedev},
to the tensor eigen problems.

The large-$N$ asymptotics in both the real and complex cases have a common expression represented concisely 
by some geometric quantities of tensors parameterized by the eigenvectors. 
The geometric structure is not evident in the mean distributions, because different  
geometric quantities cannot fully be distinguished. 
Though this paper is limited to the two kinds of symmetric tensors, 
the common expression suggests the presence of universality over
various kinds of random tensors also in the joint distributions.
This extends the universality found for the mean distributions \cite{Delporte:2025tjp}.
It would therefore be interesting future study to extend computations of 
joint distributions to other random tensors.

\vspace{.3cm}
\section*{Acknowledgements}
The author was supported in part by JSPS KAKENHI Grant No.~25K07153. 
He would like to thank  S.~Nishigaki for brief correspondence.

\appendix
\def\thesection{Appendix  \Alph{section}}

\section{Triviality of the symmetric random matrix case}
\label{app:trivial}
In this appendix we show that it is a trivial problem to study the eigenvector distributions for the symmetric
random matrices. 

Let us first consider the real case. The eigen equation is given by
\[
T_{ab} w_b=z\, w_a,
\label{eq:matrixrealsym}
\] 
where $T$ is a symmetric real matrix, $z$ the real eigenvalue, and $w$ the eigenvector. One can generally take 
the eigenvectors to be real, and can take the normalization, $|w|=1$. 

The eigen equation is invariant under the orthogonal transformation \eq{eq:rsortho}. Therefore the eigenvector
distribution for the Gaussian random tensor 
is invariant under the orthogonal transformation. By an orthogonal transformation, 
$w$ can always be transformed to $w=(1,0,\cdots,0)$. Therefore the mean distribution of the eigenvector
is trivial.

As for the joint distribution, consider any two of eigen pairs,
\[
T_{ab}w_b^{(i)} =z^{(i)} w^{(i)}_a, \ |w^{(i)}|=1, \ i=1,2.
\]
For a random matrix, we can probabilistically ignore the possibilities of degenerate eigenvalues, 
so we can assume $z^{(1)}\neq z^{(2)}$.  Then we find
\[
w_a^{(1)} T_{ab} w_b^{(2)}=z^{(1)} w_a^{(1)} w_a^{(2)}=z^{(2)} w_a^{(1)} w_a^{(2)},
\]
depending on which of $w_a^{(1)} T_{ab}$ or $T_{ab} w_b^{(2)}$ is computed first. This concludes
\[
w_a^{(1)} w_a^{(2)}=0.
\]
Therefore there are no non-trivial correlations among the eigenvectors. 

Let us now consider the complex case. The eigen equation is given by
\[
T_{ab} w^*_b=z\, w_a,
\]
where $T$ is a complex symmetric matrix. One can take the normalization $|w|=1$. 

The eigen equation is invariant under the unitary transformation \eq{eq:csunitary}.  Therefore the mean
distribution is invariant under the unitary transformation of $w$. $w$ can always be transformed to 
$w=(1,0,\cdots,0)$.  Therefore the mean distribution is trivial.

As for the joint distribution, let us take any two of eigen pairs.
\[
T_{ab} w^{(i)*}_b=z^{(i)}\, w^{(i)}_a,\ i=1,2.
\]
We can probabilistically assume $|z^{(1)}|\neq |z^{(2)}|$ for a randomly given $T$.
We obtain
\[
w^{(1)*}_a T_{ab} w^{(2)*}_b=z^{(1)} w^{(1)}_a w^{(2)*}_a=z^{(2)} w^{(1)*}_a w^{(2)}_a.
\]
By taking the absolute value of the equation,  we obtain $|w^{(1)*}_a w_a^{(2)}|=0$. Therefore there are
no correlations between the eigenvectors.

\section{Proof of commutativity of the two limits}
\label{app:commute}
In this appendix we will prove that the two limits, $\epsilon\rightarrow +0$ and $N\rightarrow \infty$, can be 
interchanged under the assumption that the $N\rightarrow \infty$ limit exists for $\epsilon=0$.  Here note that
the regularization parameter $\epsilon$ is introduced for the field theoretical computations, and is not needed
to define the distributions of the eigenvectors themselves. 
We will prove the commutativity of the two limits 
for the case of the real symmetric random tensor. The proof can straightforwardly 
be extended for the complex symmetric random tensor. 

We will define some notations which are slightly different from the main text for 
convenience. The new notations should be confined within this appendix and should not be confused 
with the ones in the other parts. Note that, in the proof, we can regard the eigenvectors as constants 
(or given values), so we will ignore them in the arguments of functions, 
while the dependences on $T$, $\epsilon$ and $N$ are explicitly shown.

Let us denote the Jacobian in \eq{eq:rsjacobian} as $J[T,N]$. 
Then the regularized determinant factor in \eq{eq:absdet} is defined by
\[
B[T,\epsilon,N]:=
\sqrt{\det \left( J[T,N]^2+\epsilon I_{N} \right) }=\prod_{i=1}^N \sqrt{e[T,N]_i^2+\epsilon},
\]
where $e[T,N]_i$ denotes the real eigenvalues of the symmetric real matrix $J[T,N]$. Here $\epsilon \geq 0$ is 
taken throughout this appendix. Note that putting $\epsilon=0$ in the above formulas is well-defined.
$B[T,\epsilon,N]$ is a continuous monotonic function of $\epsilon$ in $\epsilon \geq 0$, satisfying
\[
0\leq B[T,\epsilon_1,N]<B[T,\epsilon_2,N] \hbox{ for } 0\leq \epsilon_1 < \epsilon_2.
\label{eq:bproperty}
\]

The regularized eigenvector distribution is defined by
\[
\rho[\epsilon,N]:=\frac{1}{A}\int dT e^{-\alpha |T|^2} \, B[T,\epsilon,N] \delta[f[T,N]].
\label{eq:apprho}
\]
Note that the integrand is non-negative and it is integrable\footnote{This comes from the 
fact that $\det (J(T,N)^2+\epsilon I_N)$ is a polynomial function of $T$, and also $\delta[f[T,N]]$ gives linear 
constraints on $T$, which delete some of the components of $T$. 
Therefore $B[T,\epsilon,N] \delta[f[T,N]]$ is polynomially bounded by $|T|$.}. 
In particular, from \eq{eq:bproperty} we can apply the monotone convergence theorem:
\[
\lim_{\epsilon\rightarrow +0} \rho[\epsilon,N]= \rho[0,N].
\label{eq:apprho0}
\]

Now let us define 
\[
F[\epsilon]:=\lim_{N\rightarrow \infty} \frac{1}{N} \log \rho[\epsilon,N].
\label{eq:appdeff}
\]
The presence of this infinite limit is not a trivial matter, which requires careful computations and specification 
of scaling (as $v=N^{-1/(2(p-2))} \tv$ in Section~\ref{sec:rslargen}).
 The computations of Sections~\ref{sec:rslargen}
and \ref{sec:rslargenL} show the presence of $F[\epsilon]$ for $\epsilon>0$. One can also refer to \cite{cart},
where it has been proven that the number of eigenvalues of a certain tensor is exponential in $N$.
Note also that the regularization with $\epsilon$ has been introduced for the quantum field theoretical computations,
and the presence of $F[0]$ itself is an independent matter from the regularization. 
In this appendix, we assume the presence of $F[\epsilon]$ for $\epsilon \geq 0$.

Because of \eq{eq:bproperty} $F[\epsilon_1]$ is monotonic: 
\[
F[0]\leq F[\epsilon_1] \leq F[\epsilon_2] \hbox{ for } 0 < \epsilon_1<\epsilon_2.
\label{eq:monotonicF}
\]
Therefore there exists
\[
F^*=\lim_{\epsilon\rightarrow +0} F[\epsilon].
\label{eq:fstar}
\]

Next we want to prove $F^*=F[0]$.
Suppose it is not, $F^*>F[0]$.   Let us consider a small $\delta>0$ which satisfies $F^*>F[0]+ 2\delta$. 
From \eq{eq:monotonicF}, for any $\epsilon_c$,
\[
F^*\leq F[\epsilon] \hbox{ for } 0<\epsilon<\epsilon_c.
\label{eq:appfineq}
\]
From \eq{eq:appdeff} there also exists $N_c$ such that
\[
\left|
\frac{1}{N} \log \rho[\epsilon,N]-F[\epsilon]
\right|<\delta \hbox{ for } N>N_c.
\label{eq:applogineq}
\] 
By combining \eq{eq:appfineq} and \eq{eq:applogineq} we obtain
\[
\frac{1}{N} \log \rho[\epsilon,N]>F[0]+ \delta \hbox{ for }0<\epsilon< \epsilon_c,\ N>N_c.
\]
Then, using \eq{eq:apprho0}, we obtain
\[
\frac{1}{N} \log \rho[0,N]\geq F[0]+ \delta \hbox{ for }N>N_c.
\]
This contradicts \eq{eq:appdeff} for $\epsilon=0$. 

Therefore we conclude
\[
\lim_{\epsilon\rightarrow +0} \lim_{N\rightarrow \infty} \frac{1}{N} \log \rho(\epsilon,N)=\lim_{\epsilon\rightarrow +0} F[\epsilon]=F[0]= \lim_{N\rightarrow \infty} \lim_{\epsilon\rightarrow +0} \frac{1}{N} \log \rho(\epsilon,N).
\]

\section{Correlation functions of matrix models}
In this appendix we give the explicit forms of the correlation functions of the matrix models, which give explicit expressions of tensor eigenvector mean distributions.
The formulas are based on \cite{matrixmodel}.

\subsection{Real case}
\label{app:rsexplicitI}
The matrix model \eq{eq:rszmat} is of a Gaussian ensemble with $\beta=1$. 
In this case, the matrix model must be solved with skew orthogonal polynomials.   
We show the formulas only for $N=2 \nu$ (even numbers) for the simplicity:
\s[
&\int_0^\infty dz_1 \cdots dz_{N-1}\, P^{(N-1)}(z_1,\cdots,z_{N-1})=\frac{(2\pi)^\frac{N-1}{2}}{(N-1)!}  \prod_{j=1}^{N-1} \frac{\Gamma[1+j/2]}{\Gamma[3/2]}, \\
&\int_0^\infty dz_1 \cdots dz_{N-1} \,P^{(N)}(z_1,\cdots,z_{N})
=\frac{(2 \nu)! (\nu-1)! 2^\nu}{\nu! (2 \nu-1)!} \left( \prod_{m=0}^{\nu-1} r_m \right) S_{\nu}[z_N], 
\s]
where 
\s[
&r_m:=2^{-2m} \sqrt{\pi} \Gamma[2m+1], \\
&S_{\nu}[z]:=\sum_{m=0}^{\nu-1} \frac{1}{2 r_m} \left(
\theta_{2 m}[z] \theta'_{2m+1}[z]-\theta_{2m+1}[z] \theta'_{2m}[z]\right), \\
&\theta_m[z]:=\int_{-\infty}^z dx\, R_m[x] \exp[-x^2/2]-\frac{1}{2} \int_{-\infty}^\infty dx\, R_m[x] \exp[-x^2/2], \\
&R_m[x]:=\left\{
\begin{matrix}
&2^{-m} H_m[x] & \hbox{for }m=\hbox{even} \\
&2^{-m} H_m[x] -(m-1) 2^{-m-1} H_{m-2}[x]& \hbox{for }m=\hbox{odd}
\end{matrix}
\right.
\s]
with $H_n[x]$ being the Hermite polynomials and $'$ denoting taking a derivative.

\subsection{Complex case}
\label{app:csexplicitI}
This case corresponds to the Laguerre ensemble with $\alpha=0,\beta=1$. 
For simplicity we consider $N=2 \nu$. 
\s[
 &\int_0^\infty dz_1 \cdots dz_{N-1}\, P^{(N-1)}(z_1,\cdots,z_{N-1})=\frac{1}{(N-1)!}
 \prod_{j=0}^{2(\nu-1)} \frac{\Gamma\left[ \frac{3+j}{2}\right] \Gamma\left[1+\frac{j}{2}\right]}{\Gamma\left[ \frac{3}{2} \right]}, \\
& \int_0^\infty dz_1 \cdots dz_{N-1} \,P^{(N)}(z_1,\cdots,z_{N})=
\frac{(2\nu)! (\nu-1)! 2^\nu}{\nu!(2\nu-1)!} \left( \prod_{m=0}^{\nu-1} r_m \right) S_\nu[z_N],
\s]
where
\s[
 &r_m:=2^{-4m-2} \Gamma[2m+1]\Gamma[2m+2],\\
 &S_\nu[z]:=\sum_{m=0}^{\nu-1}\frac{1}{2 r_m}\left( \theta_{2m}[z] \theta'_{2m+1}[z] -\theta_{2m+1}[z] \theta'_{2m}[z]\right),\\
 &\theta_m[z]=\int_0^z dx\, e^{-x} R_m[x]-\frac{1}{2} \int_0^\infty dx\, e^{-x} R_m[x], \\
 &R_m[x]:=\left\{
 \begin{matrix}
 & -\frac{m!}{2^{m+1}} \frac{d}{dx}  L_{m+1}[2x] &\hbox{for }m=\hbox{even} \\
 & -\frac{m!}{2^m} L_m[2x] -\frac{m!}{2^{m+1}} \frac{d}{dx} L_{m-1}[2x] &\hbox{for }m=\hbox{odd} 
 \end{matrix}
 \right.
\s]
with $L_{m}[x]$ being the Laguerre polynomials.

\section{Form of two-field correlations}
\label{app:susytwo}
In this appendix we derive the expression \eq{eq:rsassqL} with \eq{eq:theformofQL} from \eq{eq:rssusyL}, \eq{eq:rsintphiL}, and 
the orthogonal symmetry. Note that \eq{eq:rsassq} with \eq{eq:theformofQ} is just the case with $L=1$.
\eq{eq:csassqL} can also be derived in a similar manner.

The form $I^\perp_{ab}$ on the vector space index in \eq{eq:rsassqL} and 
the form of $Q$ in \eq{eq:theformofQL} are straightforward consequences from the orthogonal symmetry 
and the exchange symmetry \eq{eq:rsintphiL}, respectively. 
Below we will study consequences of the susy transformation \eq{eq:rssusyL}.

For simplicity let us only make explicit the indices related to the susy transformation.
Using the original variables $\Phi=(\bpsi,\psi,\phi)$ in \eq{eq:detexp},
the susy transformation with $m=1$ in \eq{eq:rssusyL} is given by 
 \[
 \dels_{l1}\bpsi_{l'}=\delta_{ll'} \phi_l,\ \dels_{l1} \psi_{l'}=0,\ \dels_{l1} \phi_{l'}=\delta_{ll'} \frac{1}{2} \psi_l.
 \]
Then a susy invariance of a two-field correlation implies
\[
0=\langle \dels_{l1} (\bpsi_{l_1} \phi_{l_2})\rangle=\delta_{ll_1} \langle \phi_{l_1} \phi_{l_2} \rangle - 
\delta_{ll_2} \frac{1}{2} \langle \bpsi_{l_1} \psi_{l_2} \rangle. 
\]
Another invariance implies
\[
0=\langle \dels_{l1}( \psi_{l_1} \phi_{l_2}) \rangle=-\delta_{ll_2} \frac{1}{2} \langle \psi_{l_1}\psi_{l_2}\rangle.
\]
We obtain a similar equation for $\langle \bpsi_{l_1} \bpsi_{l_2} \rangle$ from $\dels_{l2}$. By taking
$l=l_1$ or $l=l_2$ in these equations, we conclude \eq{eq:rsassqL}.

\section{Computation of $S_{\rm eff}$} 
\label{app:SD}
Let us first show the computation of $\langle S_\perp \rangle$. 

As for one of the quadratic terms, by using \eq{eq:rsassq}, we obtain
\[
\langle G_{\kappa \kappa'} \pPhi_{\kappa}\cdot \pPhi_{\kappa'} \rangle=K_{mn} \tilde K_{ij} K^{-1}_{mn} Q_{ij} I^\perp_{aa}
=-(N-1)\tilde K_{ij}Q_{ij}=-2 (N-1) Q_2\sim -2 N Q_2,
\]
where $\sim$ denotes the leading order in $N$.
Here it was important to pay attention to the graded symmetry of $K$ (see below \eq{eq:kkt}):
$K_{mn} K^{-1}_{mn}=(-1)^{\underline{n}\, \underline{m}} K_{nm} K^{-1}_{mn}
=(-1)^{\underline{n}} K_{nm} K^{-1}_{mn}=\sum_{n=1}^3 (-1)^{\underline{n}}=-1$.
The computation of $\langle \tG_{\kappa \kappa'} \pPhi_{\kappa}\cdot \pPhi_{\kappa'} \rangle$ is similar.

As for the quadratic term,
\s[
\langle G_{\kappa \kappa'} G_{\kappa '' \kappa '''} \ss \pPhi_{\kappa} \pPhi_{\kappa'} \se \cdot 
\ss \pPhi_{\kappa''} \pPhi_{\kappa'''} \se \rangle&=\langle G_{\kappa \kappa'} G_{\kappa '' \kappa '''} 
(-1)^{\underline{\kappa'} \, \underline{\kappa''}} 
\pPhi_{\kappa}\cdot \pPhi_{\kappa''} \, \pPhi_{\kappa'} \cdot \pPhi_{\kappa'''}  \rangle \\
&\sim  G_{\kappa \kappa'} G_{\kappa '' \kappa '''} 
(-1)^{\underline{\kappa'} \, \underline{\kappa''}} 
\langle \pPhi_{\kappa}\cdot \pPhi_{\kappa''} \rangle \langle \pPhi_{\kappa'} \cdot \pPhi_{\kappa'''}  \rangle \\
&=K_{mm'} \tilde K_{ii'} K_{m''m'''} \tilde K_{i''i'''} (-1)^{m'm''} K^{-1}_{mm''} Q_{ii''} K^{-1}_{m'm'''} Q_{i'i'''} \\
&=-2(N-1)^2 (Q_1^2+Q_2^2) \\
&\sim -2N^2 (Q_1^2+Q_2^2),
\s] 
where we have paid attention to the graded symmetries of $\pPhi_\kappa$ and $K$. 
From the first to the second lines, we have ignored the lower order 
contributions.\footnote{Schematically, this can be described as
\s[
\langle \phi^{i_1}\cdot\phi^{i_2}\phi^{i_3}\cdot\phi^{i_4} \rangle&=\langle \phi^{i_1}_a\phi^{i_2}_a\phi^{i_3}_b
\phi^{i_4}_b \rangle\\ 
&= \langle \phi^{i_1}_a\phi^{i_2}_a \rangle \langle \phi^{i_3}_b
\phi^{i_4}_b \rangle+ \langle \phi^{i_1}_a\phi^{i_3}_b \rangle \langle \phi^{i_2}_a
\phi^{i_4}_b \rangle+ \langle \phi^{i_1}_a\phi^{i_4}_b \rangle \langle \phi^{i_3}_b
\phi^{i_2}_a \rangle+\langle \phi^{i_1}_a\phi^{i_2}_a \phi^{i_3}_b\phi^{i_4}_b \rangle_{\rm c} \\
&=N^2 Q_{i_1i_2} Q_{i_3 i_4}+N Q_{i_1i_3} Q_{i_2i_4}+N Q_{i_1i_4} Q_{i_3i_2}
+\langle \phi^{i_1}_a\phi^{i_2}_a \phi^{i_3}_b\phi^{i_4}_b \rangle_{\rm c} \\
&\sim N^2 Q_{i_1i_2} Q_{i_3 i_4}
\s]
for $\langle \phi^{i_1}_a\phi^{i_2}_b\rangle=Q_{i_1i_2}\delta_{ab}$,
where $\langle \cdot \rangle_{\rm c}$ denotes a connected correlation, which is assumed to be sub-leading.}

The superdeterminant term can be computed straightforwardly by 
\[
\sdet(K^{-1}_{mn} Q_{ij} I^\perp_{ab})=(\sdet(K^{-1}_{mn} Q_{ij}))^{N-1}
={\rm const.} (\det Q_{ij})^{N-1}.
\]

\section{Parameterizations without redundancies}
\label{app:volume}
Due to the orthogonal symmetry \eq{eq:rsortho} and the rotational symmetric distribution of the tensor $T$, 
the distribution of the eigenvectors $\rho_L[\{v \}]$ is invariant under the rotational transformation,
\[
v_{la}\rightarrow M_{aa'} v_{la'}, \ (l=1,2,\cdots,L),
\]
where $M$ is an arbitrary orthogonal matrix of dimension $N$. 
We can remove the redundancy of the variables in the following manner. 
Let us first consider $v_1$. By a rotational transformation
$v_1$ can be put in the form, $v_{1}=(v_{11},0,\cdots,0)\ (v_{11} > 0)$. 
Here, note that the linear independence of $v_l$ ignores $v_{11}=0$. The integration measure is transformed to
\[
dv_1'=S_{N-1}\, v_{11}^{N-1}  dv_{11}
\]
by the change of the variable. We can next use a rotational transformation within the transverse subspace
against $v_1$ to put $v_2$ in the form,
$v_{2}=(v_{21},v_{22},0,\cdots,0)\ (v_{22} > 0)$.
The integration measure is transformed to
\[
dv_2'=S_{N-2} \, v_{22}^{N-2} d v_{21} dv_{22}. 
\]
Note that $v_{22}= 0$ has been ignored by the same reason as above. 
By continuing the process (assuming $L\leq N$), we can parameterize $\{ v \}$ without redundancy in the form
\eq{eq:volgauge}, and the associated infinitesimal volume is given by \eq{eq:volfac}.

Similar procedure as above can also be carried out for the complex case by using the unitary transformation. 
A convenient parametrization is given by \eq{eq:cspara}, and the associated infinitesimal volume 
\eq{eq:csvol}.

\section{Integration over the parallel part of $\Phi$}
\label{app:rsphipara}
There exist some non-interacting degrees of freedom of $\Phi$ in the action \eq{eq:rssLp}.
To single them out, let us consider a decomposition,
\[
\Phi_{l\kappa}=\Phi^v_{l\kappa} v_l+\Phi^{v_\perp}_{l\kappa}, 
\label{eq:rsphilkdecomp}
\]
where $\Phi_{l\kappa}^{v_\perp} \cdot v_l=0\ (\hbox{no sum over }l)$.
Then we find
\s[
\tIperp\cdot \ss \G_{\kappa \kappa'} \Phi_{l\kappa} \Phi_{l\kappa'} v_l^{p-2}\se&=
\tIperp\cdot \ss
 \G_{\kappa \kappa'} \Phi_{l\kappa}^v \Phi_{l\kappa'}^v v_l^{p}
+ 2 \G_{\kappa \kappa'} \Phi_{l\kappa}^v \Phi_{l\kappa'}^{v_\perp} v_l^{p-1}
+ \G_{\kappa \kappa'} \Phi_{l\kappa}^{v_\perp} \Phi_{l\kappa'} ^{v_\perp} v_l^{p-2} \se \\
&=\tIperp\cdot \ss \G_{\kappa \kappa'} \Phi_{l\kappa}^{v_\perp} \Phi_{l\kappa'} ^{v_\perp} v_l^{p-2} \se,
\s]
because the first and the second terms on the righthand side in the first line
are projected out by $\tIperp$. Therefore $\Phi^v_{l\kappa}$ has no interactions. 
Moreover, as will be proven at the end of this appendix, 
$\Phi^v_{l\kappa}$ decouples from $\Phi^{v_\perp}_{l\kappa}$
in the second line of \eq{eq:rssLp}:
\s[
&g_{mn} \gtinv{mn}{m'n'} \ss v_{m'} v_{n'}^{p-1} \se \cdot \ss \G_{\kappa \kappa'} \Phi_{l \kappa} \Phi_{l \kappa'} v_l^{p-2} \se  \\
&=|v_l|^2 \G_{\kappa \kappa'} \Phi_{l \kappa}^v \Phi_{l \kappa'}^v
+g_{mn} \gtinv{mn}{m'n'} \ss v_{m'} v_{n'}^{p-1} \se \cdot  
\ss \G_{\kappa \kappa'} \Phi_{l \kappa}^{{v_\perp}} \Phi_{l \kappa'}^{{v_\perp}}v_l^{p-2} \se,
\label{eq:rsgginvvv}
\s]
Therefore the terms in $S_{L\Phi}$ containing $\Phi_{l\kappa}^v$ are given by
\[
S_{L\Phi}^v:=i (2-p)|v_l|^2  \G_{\kappa \kappa'} \Phi_{l \kappa}^v \cdot \Phi_{l \kappa'}^v
+\sqrt{\epsilon} |v_l|^2\tG_{\kappa \kappa'} \Phi_{l \kappa}^v \cdot \Phi_{l \kappa'} ^v,
\]
which corresponds to taking $J=|v_l|^2$ for each $\Phi_{l\kappa}^v$ in \eq{eq:detexp}. Therefore the integration over 
$\Phi^v_{l\kappa}$ 
generates a factor $(p-2)^L$ in $\epsilon\rightarrow +0$.\footnote{The effect of the coefficient $|v_l|^2$ 
cancels with the Jacobian of the integration measure of $\Phi_{l\kappa}^v$.} 
Thus we obtain
\[
Z_L=(p-2)^L Z_L^{v_\perp},
\label{eq:rspm2fac}
\]
where $Z_L^{v_\perp}$ only contains the transverse part $\Phi_{\l\kappa}^{v_\perp}$, 
\[
Z_{L}^{v_\perp}=\lim_{\epsilon\rightarrow +0} \frac{1}{\pi^{(N-1)L}}\int d\Phi^{v_\perp} \, e^{S_{L\Phi}^{v_\perp}}
\]
with
\s[
S_{L\Phi}^{v_\perp}
:=&i \G_{\kappa \kappa'} \Phi_{l \kappa}^{v_\perp}\cdot \Phi_{l \kappa'}^{v_\perp} 
+\sqrt{\epsilon} \tG_{\kappa \kappa'} \Phi_{l \kappa}^{v_\perp}\cdot \Phi_{l \kappa'}^{v_\perp}  \\
&\ 
-i (p-1) g_{mn} \gtinv{mn}{m'n'} \ss v_{m'} v_{n'}^{p-1} \se \cdot \ss \G_{\kappa \kappa'} \Phi_{l \kappa}^{v_\perp}
 \Phi_{l \kappa'}^{v_\perp} v_l^{p-2} \se \\
&\ 
-\frac{(p-1)^2}{4 \alpha}  \ss \G_{\kappa \kappa'} \Phi_{l \kappa}^{v_\perp} \Phi_{l \kappa'}^{v_\perp} v_l^{p-2} \se \cdot \tIperp \cdot
\ss G_{\kappa'' \kappa'''} \Phi_{l' \kappa''}^{v_\perp} \Phi_{l' \kappa'''}^{v_\perp} v_{l'}^{p-2} \se.
\label{eq:rssLperp}
\s]

Let us finally prove \eq{eq:rsgginvvv}.  Let us consider a decomposition,
\[
\Phi_{l\kappa}=\Phi_{l\kappa}^v v_l + \Phi_{l\kappa m}^{v_\perp V_\parallel} v_m +  \Phi_{l\kappa}^{v_\perp V_\perp},
\label{eq:rsfurtherdecomp}
\]
where $\Phi_{l\kappa}^{v_\perp}$ in \eq{eq:rsphilkdecomp} has been further decomposed into 
$\Phi_{l\kappa m}^{v_\perp V_\parallel} v_m\in V_\parallel$ and $\Phi_{l\kappa}^{v_\perp V_\perp}\in V_\perp$. 
Since $\Phi_{l\kappa}^{v_\perp}\cdot v_l=0$ (no sum over $l$), 
we have $\Phi_{l\kappa m}^{v_\perp V_\parallel} g_{lm}=0\ (\hbox{no sum over }l)$.

Let us put \eq{eq:rsfurtherdecomp} into the first line of \eq{eq:rsgginvvv}. We obtain
\s[
&g_{mn} \gtinv{mn}{m'n'} \ss v_{m'} v_{n'}^{p-1} \se \cdot \ss \G_{\kappa \kappa'} \Phi_{l \kappa} \Phi_{l \kappa'} v_l^{p-2} \se  \\
&=g_{mn} \gtinv{mn}{m'n'} \ss v_{m'} v_{n'}^{p-1} \se \cdot \Big(
 \ss v_l^{p} \se \G_{\kappa \kappa'} \Phi_{l \kappa}^v \Phi_{l \kappa'}^v
+2 \ss v_{m''} v_l^{p-1} \se \G_{\kappa \kappa'} \Phi_{l \kappa}^v \Phi_{l \kappa' m''}^{v_\perp V_\parallel}\\
&\hspace{5cm}+
\ss v_{m''}v_{m'''} v_l^{p-2} \se \G_{\kappa \kappa'} \Phi_{l \kappa m''}^{v_\perp V_\parallel} \Phi_{l \kappa' m'''}^{v_\perp V_\parallel}
\Big)\\
&=|v_l|^2 \G_{\kappa \kappa'} \Phi_{l \kappa}^v \Phi_{l \kappa'}^v
+g_{mn} \gtinv{mn}{m'n'} \ss v_{m'} v_{n'}^{p-1} \se \cdot  
\ss v_{m''}v_{m'''} v_l^{p-2} \se \G_{\kappa \kappa'} \Phi_{l \kappa m''}^{v_\perp V_\parallel} \Phi_{l \kappa' m'''}^{v_\perp V_\parallel},
\label{eq:rsshow1}
\s]
where we have used (with no sum over $l$)
\s[
&g_{mn} \gtinv{mn}{m'n'} \ss v_{m'} v_{n'}^{p-1} \se \cdot \ss v_l^{p} \se =g_{mn} \gtinv{mn}{m'n'} \gt{m'n'}{ll}=g_{ll}=|v_l|^2, \\
&g_{mn} \gtinv{mn}{m'n'} \ss v_{m'} v_{n'}^{p-1} \se \cdot \ss v_{m''} v_l^{p-1} \se \Phi_{l \kappa' m''}^{v_\perp V_\parallel}=g_{mn} \gtinv{mn}{m'n'}  \gt{m' n'}{m''l}   \Phi_{l \kappa' m''}^{v_\perp V_\parallel} =g_{ml}  \Phi_{l \kappa' m}^{v_\perp V_\parallel}=0, 
\s]
and $\Phi_{l\kappa}^{v_\perp V_\perp}\cdot v_m=0$ for ${}^\forall v_m\in V_\parallel$.
Since $\Phi_{l\kappa}^{v_\perp V_\perp}\cdot v_m=0$, \eq{eq:rsshow1} implies \eq{eq:rsgginvvv}.

\section{Expression of $|\det \Jc|$}
\label{app:csdetj}
In this appendix we derive \eq{eq:csdetjc}, the quantum field theoretical expression of $| \det \Jc |$ for the complex case. The main difference from the 
real case is that it is convenient to consider a combined component $(x^*,x) \in V^*\oplus V$ 
rather than a single variable $x$ to 
obtain expressions parallel to the real case. 

Let us start with that an integral over a complex variable $\phi$ satisfies 
\[
\frac{1}{\pi}\int d\phi \, \exp\left[- (\phi \ \phi^*) A \svec{\phi^*}{\phi}\right]=\frac{1}{\sqrt{\det A}},
\label{eq:appbos}
\]
where $A$ is a 2-dimensional complex matrix chosen so that the integrand damps exponentially at $|\phi|\rightarrow \infty$.
One can prove this by performing a unitary transformation $(\phi^*\  \phi)=M\svec{\phi_{\rm R}}{\phi_{\rm I}}$ 
with a unitary matrix $M$, which transforms the integral to an integral over real variables, 
$\int d\phi_{\rm R} d\phi_{\rm I}\,\exp\left[ -(\phi_{\rm R}\ \phi_{\rm I}) (M^\dagger A M) \svec{\phi_{\rm R}}{\phi_{\rm I}} \right]=\pi/\sqrt{\det A }$. 

By using \eq{eq:appbos}
and the same trick as \eq{eq:detexp}, we find 
\[
|\det \Jc|=\lim_{\epsilon\rightarrow +0} \frac{1}{\pi^{2N}} \int  d\phi d\bpsi d\psi \,\exp[s],
\]
where 
\[
s=(\bpsi_1,\bpsi^*_1,\bpsi_2,\bpsi^*_2)\cdot \tilde J^{\rm c} \cdot \left( \begin{matrix}  \psi_1^* \\ \psi_1 \\ \psi^*_2 \\ \psi_2 \end{matrix} \right)
-(\phi_1,\phi^*_1,\phi_2,\phi^*_2)\cdot \tilde J^{\rm c} \cdot \left( \begin{matrix}  \phi^*_1 \\ \phi_1 \\ \phi^*_2 \\ \phi_2 \end{matrix} \right)
\label{eq:apps}
\]
with
\[
\tilde J^{\rm c}:=\left( 
\begin{matrix}
\sqrt{\epsilon} I_{2N} & i J^{\rm c} \\
i J^{\rm c} &  \sqrt{\epsilon} I_{2N}
\end{matrix}
\right).
\]

It is also convenient to introduce a superfield for the complex case. 
An obstacle in doing so in parallel with the real case is that the bosonic fields in \eq{eq:apps} appear asymmetrically 
between the both sides. To make it symmetric, we introduce a matrix to do a twist,
\[
\bk:=
\left(
\begin{matrix}
0 & I_N \\
I_N & 0
\end{matrix}
\right),
\label{eq:kbar}
\]
to interchange the original and the complex conjugates. 
Then we see that 
\[
s=(\bpsi^*_1,\bpsi_1,\bpsi^*_2,\bpsi_2)\cdot \tilde J^{\rm c}{}' \cdot \left( \begin{matrix}  \psi_1^* \\ \psi_1 \\ \psi^*_2 \\ \psi_2 \end{matrix} \right)
-(\phi_1^*,\phi_1,\phi_2^*,\phi_2)\cdot \tilde J^{\rm c}{}' \cdot \left( \begin{matrix}  \phi^*_1 \\ \phi_1 \\ \phi^*_2 \\ \phi_2 \end{matrix} \right),
\label{eq:csrewrites}
\]
where 
\[
\tilde J^{\rm c}{}'=\left( 
\begin{matrix}
\sqrt{\epsilon} \bk & i \bk\star  J^{\rm c}{} \\
i \bk\star J^{\rm c}{} &  \sqrt{\epsilon} \bk
\end{matrix}
\right).
\]
Note that $\bk \star J^{\rm c}{}$ is symmetric in the current case, as shown explicitly in \eq{eq:cskj}. Therefore \eq{eq:csrewrites} is simply the same as  
the real case \eq{eq:detexp} with the replacement, 
$\bpsi_i \rightarrow (\bpsi_i^*,\bpsi_i)$, $\psi_i \rightarrow (\psi_i^*,\psi_i)$, $\phi_i\rightarrow (\phi^*_i,\phi_i)$.

\section{Explicit forms of $S_{L\Phi M}$ and $J_l(M,\{v\})$}
\label{app:actualnumerics}
\subsection{Real case}
\label{app:explicit}
In this appendix we obtain more explicit expressions of $S_{L\Phi M}$ in \eq{eq:rsslpm} and $J_l(M,\{v\})$ 
in \eq{eq:rsLmat}. 
For convenience we decompose  the components of $\Phi_{l\kappa}$ and $M_l$ according to the vector space decomposition 
$V=V_\parallel\oplus V_\perp$, where the former is expanded in terms of $v_l \ (l=1,2,\cdots,L)$. 
Namely, we decompose $\Phi_{l\kappa}$ into $\hp_{l\kappa} \in V_\perp$ and 
$\hp_{l\kappa m} v_m \in V_\parallel$, and similarly, $M_l$ into $\hm_{l}\in V_\perp\otimes V_\perp$, 
$\hm_{lm} v_{m} \in  V_\parallel\otimes V_\perp $, and $\hm_{lmn} \ss v_{m} v_{n} \se 
 \in V_\parallel \otimes V_\parallel$. 
Since $M_l$ is symmetric, $\hm_l$ is also symmetric and $\hm_{lmn}=\hm_{lnm}$. 
We specifically use $a,b$ for the indices belonging to $V_\perp$ in this appendix. 
Namely, for instance, the components of $\hm_{l}$
are represented by $\hm_{lab}$. 

Let us first discuss the quadratic term of $M_l$ in \eq{eq:rsslpm}. Since $V_\perp$ is conserved by $\tIperp$, it
generally has the form,
\[
\ss M_l v_l^{p-2} \se \cdot \tIperp \cdot \ss M_{l'} v_{l'}^{p-2} \se = \K_{ll'} \hm_{lab} \hm_{l'ab} +
\K_{lm\, l'm'} \hm_{lma} \hm_{l'm'a} + \K_{lmn\, l'm'n'} \hm_{lmn} \hm_{l'm'n'}.
\label{eq:defofK}
\] 
By explicitly putting the decomposition of $M_l$ on the lefthand side, we obtain
\s[
&\K_{ll'}=\frac{2}{p(p-1)} \gq{p-2}{l}{l'}, \\
&\K_{lm\, l'm'}=\frac{1}{p} \left( \gtwo{1}{p-2}{ml}{m'l'}-\gtwo{1}{p-2}{ml}{ii} \gqinv{p-1}{i}{i'}
\gtwo{1}{p-2}{i'i'}{m'l'} \right), \\
&\K_{lmn\,l'm'n'}=\gthree{1}{1}{p-2}{mnl}{m'n'l'}-\gthree{1}{1}{p-2}{mnl}{ijj} \gtinv{ij}{i'j'} \gthree{1}{1}{p-2}{i'j'j'}{m'n'l'},
\label{eq:rsdefofK}
\s]
where
\[
&\gtwo{q_1}{q_2}{mn}{m'n'}=\ss v_m^{q_1} v_n^{q_2} \se \cdot \ss v_{m'}^{q_1} v_{n'}^{q_2}  \se,
&\gthree{q_1}{q_2}{q_3}{mnl}{m'n'l'}=\ss v_m^{q_1} v_n^{q_2}v_l^{q_3} \se \cdot \ss v_{m'}^{q_1} v_{n'}^{q_2}v_{l'}^{q_3} \se,
\]
which extend the notations in Section~\ref{sec:rshighernot}.  With the same coefficients we also have 
\s[
&\ss M_l v_l^{p-2} \se \cdot \tIperp \cdot \ss \Phi_{l'\kappa} \Phi_{l'\kappa'} v_{l'}^{p-2} \se \\
&\ \ 
= \K_{ll'} \hm_{lab} \G_{\kappa \kappa'} \hp_{l'\kappa a} \hp_{l'\kappa'b} +
2 \K_{lm\, l'm'} \hm_{lma}  \G_{\kappa \kappa'}\hp_{l'\kappa m'} \hp_{l'\kappa'a}
 + \K_{lmn\, l'm'n'}  \G_{\kappa \kappa'} \hp_{l'\kappa m n} \hp_{l'\kappa' m' n'} .
\s] 

\eq{eq:rsslpm} also contains a term,
$
Y \cdot \ss \G_{\kappa \kappa'} \Phi_{l \kappa}^{} \Phi_{l \kappa'}^{} v_l^{p-2} \se=Y_{lmn} G_{\kappa \kappa'} \Phi_{l \kappa m}^{} \Phi_{l \kappa'n}^{},
$
where
\s[
Y_{lmn}:=g^{(1,1,p-2)}_{mnl\,ijj} g^{(1,p-1)\,-1}_{ij\,i'j'} g_{i'j'}.
\label{eq:rsdefofL}
\s]

The expressions above will give the explicit form of $S_{L\Phi M}$, but it is inconvenient for the numerical study because of the 
mixture of the components of $\hm$. 
The coefficients $\K_{ll'}$, $\K_{lm\, l'm'}$, and  $\K_{lmn\, l'm'n'}$ can be regarded as real symmetric matrices 
in terms of the former and latter index sets, namely $l$ and $l'$, $lm$ and $l'm'$, and $lmn$ and  $l'm'n'$, respectively. 
Therefore they can be diagonalized by orthogonal matrices, $\A_{ll'}$, $\A_{lm\, l'm'}$, and  $\A_{lmn\, l'm'n'}$, respectively:
\s[ 
&\K_{ll'}=\A_{ll''} \A_{l'l''} \e_{l''}, \\
&\K_{lm\,l'm'}=\A_{lm \, l''m''} \A_{l'm'\,l''m''} \e_{l''m''}, \\
&\K_{lmn\,l'm'n'}=\A_{lmn \, l''m''n''} \A_{l'm'n'\,l''m''n''} \e_{l''m''n''},
\s]
where $\e_{l''},\e_{l''m''},\e_{l''m''n''}$ are the eigenvalues of the $\K$'s. We also perform the diagonalization of the 
kinetic term of $\Phi_{l\kappa}$ by introducing a diagonalization $g_{mm'}=\B_{mm''} \B_{m'm''} \ep_{m''}$ 
with an orthogonal matrix $\B$. Then we obtain
\s[
S_{L\tP\tM}:=& i \G_{\kappa \kappa'} \left( \tP_{l\kappa m} \tP_{l\kappa' m} +\tP_{l\kappa a}\tP_{l\kappa' a} \right)
+\sqrt{\epsilon} \tG_{\kappa  \kappa'} 
\left( \tP_{l\kappa m} \tP_{l\kappa' m} +\tP_{l\kappa a}\tP_{l\kappa' a} \right)\\
&-i(p-1) \tilde Y_{lmn} \G_{\kappa\kappa'} \tP_{l\kappa m}  \tP_{l\kappa'n}
-\frac{i(p-1)}{\sqrt{\alpha}}\Big( \tK_{ll'} \tM_{lab} \G_{\kappa\kappa'} \tP_{l'\kappa a}\tP_{l'\kappa'b} \\
&+2 \tK_{lm\,l'm'}\tM_{lma}\G_{\kappa\kappa'} \tP_{l'\kappa m'}\tP_{l' \kappa' a}
+\tK_{lmn\,l'm'n'} \tM_{lmn}\G_{\kappa\kappa'} \tP_{l'\kappa m'} \tP_{l'\kappa' n'}  
\Big)\\
&-\tM_{lab} \tM_{lab}-\tM_{lma}\tM_{lma}-\tM_{lmn}\tM_{lmn}, 
\label{eq:rsslp}
\s]
where
\s[
&\tK_{ll'}:=\A_{l'l} \sqrt{\e_l},\\
&\tK_{lm\, l'm'}:=\A_{l'm''\, lm} \sqrt{\e_{lm}} \B_{m''m'}\frac{1}{\sqrt{\ep_{m'}}}, \\
&\tK_{lmn\,l'm'n'}:=\A_{l'm''n''\, lmn} \sqrt{\e_{lmn}} \B_{m''m'} \frac{1}{\sqrt{\ep_{m'}}} \B_{n''n'} \frac{1}{\sqrt{\ep_{n'}}}, \\
&\tilde Y_{lmn}:=g_{n''m''} \gtinv{m''n''}{m'''n'''} \gthree{1}{1}{p-2}{m'''n'''n'''}{m'n'l} \B_{m'm} \frac{1}{\sqrt{\ep_{m}}} \B_{n'n}
\frac{1}{\sqrt{\ep_{n}}}.
\label{eq:rstk}
\s]
Here we have introduced new variables $\tP$ and $\tM$ after transforming $\Phi$ and $M$ by
the orthogonal transformations and rescaling them by the eigenvalues so that some of their quadratic terms be
normalized. Note that, since we assume $v_m$ are linearly independent, the eigenvalues $\ep_m$ are 
positive, making \eq{eq:rstk} well defined. On the other hand, the eigenvalues $e_l,e_{lm},e_{lmn}$ 
are non-negative\footnote{Because of the non-negative definition \eq{eq:defofK}.} 
and can be zeros, which corresponds to the presence of the kernel \eq{eq:rskernel}.
Note also that there appears no Jacobian in the transformation from $\Phi$ to $\tP$, while 
the Jacobian from $M$ to $\tM$ does not matter because of the cancellation between 
the denominator and the numerator in \eq{eq:rszltilde} below.

By using the above parametrizations and \eq{eq:detexp}, we obtain
\[
Z_L=\frac{1}{Z_L^0}\int_{V_{\tM}} d\tM\, \prod_{l=1}^L \left| \det \tJ_l  \right| \, \exp\left[ -\tM_{lab} \tM_{lab}-\tM_{lma}\tM_{lma}-\tM_{lmn}\tM_{lmn} \right],
\label{eq:rszltilde}
\]
where 
\[
&Z_L^0 :=\int_{V_{\tM}} d\tM \exp\left[ -\tM_{lab} \tM_{lab}-\tM_{lma}\tM_{lma}-\tM_{lmn}\tM_{lmn} \right], \\
&\tJ_l:=\left(
\begin{matrix}
\tJ^{(l1)} & \tJ^{(l2)} \\
\tJ^{(l3)} &\tJ^{(l4)} 
\end{matrix}
\right)
\]
with the sub-matrices having components,
\s[
\tJ^{(l1)}_{mn}:=&\delta_{mn}-(p-1) \tilde Y_{lmn} -\frac{p-1}{\sqrt{\alpha}} \tM_{l'm'n'} \tK_{l'm'n'\,lmn},\\
\tJ^{(l2)}_{ma}=&\tJ^{(l3)}_{am}:=-\frac{p-1}{\sqrt{\alpha}}\tM_{l'm'a} \tK_{l'm'\,lm},\\
\tJ^{(l4)}_{ab}:=&\delta_{ab}-\frac{p-1}{\sqrt{\alpha}} \tM_{l'ab} \tK_{l'l}.
\s]
Here it should be reminded that the indices $a,b$ run only through $N-L$ dimensions. 

The integration region $V_{\tM}$ in \eq{eq:rszltilde} is simply the whole space of real matrices 
(with symmetries, $M_{lab}=M_{lba}$ and $M_{lmn}=M_{lnm}$), which is in contrast with \eq{eq:defofvm}, where a kernel must be modded out. 
This modding-out is automatic in \eq{eq:rszltilde}, because the kernel corresponds to the zero eigenvalue spaces of $\e_l,\e_{lm},\e_{lmn}$ and 
its contribution automatically drops out from $\tJ_l$ due to \eq{eq:rstk}.

\subsection{Complex case}
\label{app:csexplicit}
In this appendix, we obtain the explicit formulas for the complex case $\ZcL$ in \eq{eq:csLmat}. The expressions can straightforwardly be obtained by 
using the same definitions of $K$ in \eq{eq:rsdefofK} and $Y$ in \eq{eq:rsdefofL} with the replacement 
of the metrics with the complex ones, 
$g^{(q_1,\cdots,q_s)}_{m_1\cdots m_s\, n_1\cdots n_s}=\ss v_{m_1}^{*q_1}\cdots v_{m_s}^{* q_s} \se\cdot \ss v_{n_1}^{q_1}\cdots v_{n_s}^{q_s} \se$. 
In the complex case $K$'s are hermitian matrices, and can be diagonalized by unitary matrices $A$, 
as $K_{ll'}=A_{ll''}A^*_{l'l''} e_{l''}$, and so on. 
Similarly to \eq{eq:rstk}, let us accordingly define
\s[
&\tilde K_{ll'}:=A^*_{l'l} \sqrt{e_l},\\
&\tilde K_{lm\,l'm'}:=A^*_{l'm''\,lm} \sqrt{e_{lm}} B_{m'' m'}\frac{1}{\sqrt{\ep_{m'}}}, \\
&\tilde K_{lmn\,l'm'n'}:=A^*_{l'm''n''\,lmn} \sqrt{e_{lmn}} B_{m'' m'}\frac{1}{\sqrt{\ep_{m'}}}B_{n'' n'}\frac{1}{\sqrt{\ep_{n'}}},\\
&\tilde Y^*_{lmn}:=g_{n''m''} \gtinv{m''n''}{m'''n'''} \gthree{1}{1}{p-2}{m'''n'''n'''}{m'n'l} \B_{m'm} \frac{1}{\sqrt{\ep_{m}}} \B_{n'n}\frac{1}{\sqrt{\ep_{n}}}.
\s]
Then, by performing a similar manipulation, we obtain  
\[
\ZcL=\frac{1}{\ZcL{}^0} \int_{V^{\rm c}_{\tM}} d\tM\, \prod_{l=1}^L \left| \det \tJ_l^{\rm c}  \right| \, \exp\left[ -\tM_{lab}^* \tM_{lab}-\tM_{lma}^* \tM_{lma}
-\tM_{lmn}^*\tM_{lmn} \right],
\label{eq:csrszltilde}
\]
where
\[
\tilde J^{\rm c}_l:=\mat2{I_N & \tilde J'_l}{\tilde J'^*_l & I_N}
\]
with the $N$-dimensional sub-matrix defined by 
\s[
&\tilde J'_l:=\mat2{\tilde J'^{(l1)} & \tilde J'^{(l2)}}{\tilde J'^{(l3)} & \tilde J'^{(l4)}}, \\
&\tilde J'^{(l1)}_{mn}:=-(p-1) \tilde Y^*_{lmn}-\frac{p-1}{\sqrt{\alpha}} \tilde M^*_{l'm'n'} \tilde K _{l'm'n'\,lmn},\\
&\tilde J'^{(l2)}_{ma}=\tilde J'^{(l3)}_{am}:=-\frac{p-1}{\sqrt{\alpha}}\tilde M^*_{l'm'a} \tilde K _{l'm'\,lm},\\
&\tilde J'^{(l4)}_{ab}:=-\frac{p-1}{\sqrt{\alpha}}\tilde M^*_{l'ab} \tilde K _{l'l}.\\
\s]

\end{document}